\documentclass[aps,prx,12pt,superscriptaddress]{revtex4-2}
\usepackage{graphicx}
\usepackage{bm}
\usepackage{mathrsfs}
\usepackage{subcaption}
\usepackage{amsmath}
\usepackage{amsfonts}
\newcommand\vareps{\varepsilon}

\renewcommand\d\partial

\newcommand\<\langle
\renewcommand\>\rangle

\usepackage[colorlinks=true,citecolor=blue,urlcolor=blue,linkcolor=black,pdfstartview=FitH,bookmarksopen]{hyperref}
\newcommand{\be}{\begin{equation}}
\newcommand{\ee}{\end{equation}}
\newcommand{\bea}{\begin{eqnarray}}
\newcommand{\eea}{\end{eqnarray}}
\newcommand{\ie}{\begin{equation}\begin{aligned}}
\newcommand{\fe}{\end{aligned}\end{equation}}

\newcommand\p{\mathbf{p}}

\renewcommand\r{\mathbf{r}}

\usepackage{xcolor}
\usepackage{amssymb} 
\usepackage{comment}
\DeclareMathOperator{\Tr}{Tr}

\begin{document}
\title{Controlled Theory of Skyrmion Chern Bands in Moir\'e Quantum Materials: Quantum Geometry and Collective Dynamics}

\author{Yi-Hsien Du}
\email[Electronic address:$~~$]{yhdu@mit.edu}
\affiliation{Department of Physics, Massachusetts Institute of Technology, Massachusetts 02139, USA}

\begin{abstract}
Recent experiments in moir\'e quantum materials exhibit quantized Hall states without an external magnetic field, motivating continuum mechanisms based on smooth moir\'e-periodic pseudospin textures. We present a controlled theory of skyrmion Chern bands generated by such textures. An exact local $SU(2)$ transformation reveals an emergent non-Abelian gauge field; for large branch splitting we perform an operator-level Schrieffer-Wolff expansion, yielding a single-branch Hamiltonian together with systematically dressed physical operators that define the projected interacting theory beyond strict adiabaticity. The leading dynamics is governed by a $U(1)$ Berry connection whose flux is set by the skyrmion density, while controlled non-adiabatic corrections are fixed by the texture's real-space quantum geometric tensor. In a Landau-level representation built from the averaged emergent field, moir\'e-periodic modulations induce Umklapp-resolved deformations of Girvin-MacDonald-Platzman kinematics and microscopic sources of excess optical quantum weight above the topological lower bound. Assuming a gapped Hall phase, we further derive a skyrmion-crystal effective field theory with a universal Berry-phase term and a noncommutative magnetophonon. Our results provide experimentally accessible signatures for twisted transition-metal dichalcogenide homobilayers and rhombohedral graphene aligned with hexagonal boron nitride.
\end{abstract}

\maketitle
\tableofcontents

\section{Introduction}

Moir\'e superlattices have turned van der Waals heterostructures into tunable platforms for correlation-driven topological phases. In several settings, narrow and isolated Chern bands emerge at zero external magnetic field, and interactions stabilize integer and fractional quantized Hall states. Zero-field quantum Hall plateaus have been observed in twisted transition-metal dichalcogenide (TMD) homobilayers and in graphene-based heterostructures, establishing moir\'e Chern bands as a route to quantum anomalous Hall phases and fractional Chern insulators (FCIs)~\cite{Cai_2023,Park_2023,Zeng_2023,Lu_2024,Xie_2021}. Graphene aligned with hexagonal boron nitride (hBN) provides an additional knob: stacking and alignment can strongly reshape Berry curvature and topology in moir\'e minibands~\cite{SongSamutpraphootLevitov_PNAS2015}. In parallel, field-induced fractional quantum Hall states have been observed in systems with strong superlattice potentials~\cite{Spanton_2018,aronson2024displacementfieldcontrolledfractionalchern}. At compressible fractional fillings---notably near an effective half filling in an emergent Landau-level description---long-wavelength composite-fermion theories based on flux attachment provide a complementary framework~\cite{HLR1993,Son2015DiracCF}. Optical probes such as terahertz (THz) spectroscopy and inelastic Raman scattering give direct access to neutral collective modes~\cite{paul2025shininglightcollectivemodes,kousa2025theorymagnetorotonbandsmoire,du2025chiralgravitontheoryfractional}.

Despite their close connection to quantum Hall physics, moir\'e Chern bands are not simply continuum Landau levels in disguise. Berry curvature and quantum metric are generally inhomogeneous, and the projected density operators carry intrinsic Umklapp structure: density components at momenta differing by moir\'e reciprocal lattice vectors are coupled. These microscopic features can reorganize collective spectra and redistribute oscillator strength, making finite-crystal-momentum modes visible in nominally long-wavelength channels through Umklapp-induced mode folding. A theory that keeps both topology and moir\'e periodicity explicit is therefore essential for interpreting optical spectroscopy and for diagnosing how closely a given Chern band approaches the ideal Landau-level limit.

A particularly transparent route to Chern bands in continuum moir\'e models is the formation of smooth periodic pseudospin textures. Prior analyses in twisted TMD homobilayers and rhombohedral graphene aligned with hBN settings indicate that interactions can stabilize moir\'e-periodic layer/pseudospin textures carrying nonzero skyrmion winding per moir\'e unit cell. When electrons locally polarize along a slowly varying texture, adiabatic motion in the low-energy branch acquires an emergent real-space Berry connection whose curl equals the skyrmion density. The associated emergent flux is quantized per unit cell yet generally inhomogeneous, linking real-space texture geometry to band topology and measurable charge response.

A systematic theory of this setting must address three intertwined complications. First, the adiabatic mapping is only approximate in realistic moir\'e systems: the exchange splitting $J$ that locks pseudospin to the texture is large but finite, so non-adiabatic branch mixing must be organized systematically rather than assumed away. Second, moir\'e periodicity enforces Umklapp structure in projected operators; consequently, collective modes at finite crystal momentum can transfer spectral weight into the uniform channel through mode folding. Third, when a periodic skyrmion texture forms on top of a gapped Hall background, the relevant low-energy degrees of freedom are lattice deformations and defects of a topological crystal; their parity-odd kinematics should be fixed by quantized Hall response, not by a purely local gradient expansion of a ferromagnetic sigma model.

In this work, we develop a largely analytical framework that resolves these issues in a controlled regime. The organizing principle is an operator-level projection for two pseudospin branches separated by a large local splitting $J(\bm r)$ whose instantaneous eigenbasis varies slowly in space. We first implement an exact local $SU(2)$ rotation into the texture-aligned frame, exposing an emergent non-Abelian gauge field. We then perform an operator-level Schrieffer-Wolff (SW) expansion in $1/J$ to integrate out the high-energy branch, yielding an effective single-branch Hamiltonian and systematically dressed physical operators (notably the density operator) that together define the interacting projected theory beyond strict adiabaticity. Next we reorganize the single-branch dynamics as Landau levels in the spatially averaged emergent field, perturbed by moir\'e-periodic magnetic and scalar harmonics; this uniform-plus-periodic representation keeps Umklapp mixing explicit and provides a practical starting point for response calculations. Finally, when deriving the skyrmion-crystal effective field theory (EFT), we assume the projected interacting system forms a fully gapped Hall phase so that its long-wavelength parity-odd response is captured by a Chern-Simons term with quantized coefficient $\kappa$.

Our main results are:

\emph{Single-branch reduction and operator mapping beyond strict adiabaticity.}
An operator-level SW expansion yields a gauge-covariant single-branch Hamiltonian and consistently dressed observables (density, current, etc.). The leading term reproduces the emergent $U(1)$ Berry connection, while the leading corrections are fixed by the texture's real-space quantum geometric tensor. This provides a controlled microscopic starting point for the interacting projected problem.

\emph{Uniform-plus-periodic Landau-level representation and deformed Girvin-MacDonald-Platzman kinematics.}
In the averaged emergent field, moir\'e-periodic magnetic/scalar modulations and the leading $1/J$ terms produce an Umklapp-resolved deformation of Girvin-MacDonald-Platzman (GMP) kinematics~\cite{PhysRevB.33.2481}, providing a controlled description of departures from ideal Landau-level algebra in skyrmion Chern bands.

\emph{Quantum-geometry diagnostics and the topological quantum-weight bound.}
For any gapped Hall phase, the long-wavelength longitudinal optical weight obeys a topological lower bound~\cite{PhysRevX.14.011052}. Within the texture-induced mapping we identify two controlled microscopic sources of excess weight above this minimum: Berry-curvature harmonics (flux inhomogeneity) and finite-$J$ inter-branch mixing encoded by SW corrections and operator dressing.

\emph{Umklapp-resolved response and mode folding.}
We formulate density form factors in a magnetic Bloch basis and derive an Umklapp-resolved matrix-response formalism. A robust consequence is mode folding: modes centered at finite Umklapp momentum generically acquire optical weight in the uniform channel, making them accessible in long-wavelength probes such as THz spectroscopy and Raman scattering.

\emph{Skyrmion-crystal effective field theory from a gapped Hall background.}
Assuming a gapped Hall phase, integrating out the electrons yields a Chern-Simons term for the combined gauge field. Expanding about a periodic skyrmion-crystal saddle point produces a magneto-elastic EFT with a universal Berry-phase term fixed by the Hall coefficient, implying a noncommutative magnetophonon. For short-range elasticity, the magnetophonon dispersion relation satisfies $\omega\sim q^2$, while long-range Coulomb interactions produce a crossover to $\omega\sim q^{3/2}$; weak moir\'e pinning yields a pinned resonance.

\paragraph*{Scope.} Our controlled results concern the operator-level reduction to a single low-energy branch in powers of $1/J$, yielding an effective one-branch Hamiltonian and systematically dressed observables. This construction does not by itself determine which correlated phase is realized at a given filling; rather, it provides a controlled microscopic starting point for many-body physics after projecting interactions into the low-energy branch. When deriving the skyrmion-crystal effective field theory we assume the projected interacting system forms a gapped Hall phase, so that its long-wavelength parity-odd response is captured by a Chern-Simons term with quantized coefficient $\kappa$. Here ``gapped Hall phase'' refers to a bulk electronic charge gap; the skyrmion crystal still supports gapless phonons in the absence of pinning

The paper is organized as follows. Sec.~\ref{sec:continuum_model} introduces the continuum model and the texture-induced emergent-field mapping. Sec.~\ref{sec:controlled_expansions} develops the controlled beyond-adiabatic $1/J$ expansion and the uniform-plus-periodic Landau-level representation. Secs.~\ref{sec:noncommutative} and~\ref{sec:magnetic_bloch} analyze long-wavelength kinematics and Umklapp-resolved response. Secs.~\ref{sec:effective_field_theory} and~\ref{sec:noncommutative_magnetic_translations} derive the skyrmion-crystal effective field theory and its noncommutative phonon kinematics. 

\section{Continuum Moir\'e Model and Skyrmion Chern-Band Mapping}
\label{sec:continuum_model}

We start from an interacting continuum description appropriate for small-angle moir\'e heterostructures. Throughout we often focus on a single time-reversal sector, e.g. the $K$ valley of valence-band holes, which is spin polarized by spin-valley locking, and suppress the corresponding label. When needed, the time-reversal partner sector is included explicitly. This section defines the standard continuum moir\'e single-particle Hamiltonian and fixes notation; interactions enter the later analysis through a controlled projection onto a low-energy branch and the corresponding dressed density operators (Sec.~\ref{sec:controlled_expansions}).

\subsection{Continuum model for twisted transition-metal dichalcogenide homobilayers}

In group-VI twisted transition-metal dichalcogenide homobilayers, strong spin-orbit coupling leads to spin-valley locking: within a fixed valley the relevant valence-band holes are effectively spin polarized, while time reversal interchanges valleys and flips spin. For an AA-aligned homobilayer twisted by a small angle, the single-valley continuum theory reduces to a two-component layer pseudospin. We write the continuum single-particle Hamiltonian~\cite{PhysRevLett.122.086402} as
\begin{equation}
H_0=\sum_{s=\uparrow,\downarrow}\int d^2r\;
\psi^\dagger_{s}(\bm r)\mathcal H_{s}(\bm r)\psi_{s}(\bm r)~,
\end{equation}
where $\psi_s=(\psi_{t,s},\psi_{b,s})^T$ is a layer spinor and $\mathcal H_{\downarrow}$ is the time-reversal conjugate of $\mathcal H_{\uparrow}$. In what follows we present $\mathcal H_{\uparrow}$ explicitly; the spin $s=\downarrow$ sector follows by time reversal. For spin $s=\uparrow$ we use the standard continuum Hamiltonian
\begin{equation}\label{eq:continuum}
\mathcal{H}_{\uparrow}(\bm r)=
\begin{pmatrix}
\frac{(- i \nabla - \bm\kappa_+)^2}{2m} + V_1(\bm r) & T(\bm r) \\
T^{\dagger}(\bm r) & \frac{(- i\nabla - \bm\kappa_-)^2}{2m} + V_2(\bm r)
\end{pmatrix},
\end{equation}
with effective mass $m$, smooth moir\'e-periodic intralayer potentials $V_{1,2}(\bm r)$, and moir\'e-periodic interlayer tunneling $T(\bm r)$. The vectors $\bm\kappa_{\pm}$ encode the folding of the monolayer valley momenta into the moir\'e Brillouin zone (top: $\bm\kappa_t\equiv\bm\kappa_+$; bottom: $\bm\kappa_b\equiv\bm\kappa_-$).

\paragraph*{Many-body Hamiltonian.} Electron-electron interactions are encoded in
\begin{equation}\label{eq:HC_def}
H = H_0 + H_C~, \qquad H_C=\frac{1}{2}\int d^2r\,d^2r'\;V_C(\bm r-\bm r'):\rho(\bm r)\rho(\bm r'):~,
\end{equation}
with $\rho(\bm r)\equiv \sum_{s}\psi^\dagger_{s}(\bm r)\psi_{s}(\bm r)$. 
In the remainder of the paper we first organize the single-particle moir\'e problem and then develop a controlled single-branch reduction in powers of $1/J$; the projected interacting theory is obtained by expressing $H_C$ in terms of the corresponding dressed density operator in the low-energy branch.

\paragraph*{Moir\'e reciprocal lattice.} We define the moir\'e reciprocal vectors
\begin{equation}
\mathbf{g}_i=\frac{4\pi}{\sqrt{3}a_M}\left(\cos\frac{\pi(i-1)}{3},\,\sin\frac{\pi(i-1)}{3}\right)~, \qquad i=1,\dots,6~,
\end{equation}
where the moir\'e period is $a_M=\frac{a_0}{2\sin(\theta/2)}\approx a_0/\theta$ at small twist angle $\theta$ (monolayer lattice constant $a_0$). A convenient choice of folded valley momenta is
\begin{equation}
\bm\kappa_-=\frac{\mathbf{g}_1+\mathbf{g}_6}{3}~,\qquad \bm\kappa_+=\frac{\mathbf{g}_1+\mathbf{g}_2}{3}~,
\end{equation}
noting that alternative conventions are related by moir\'e reciprocal-lattice shifts and/or a shift of the real-space origin.

\subsection{Lowest-harmonic moir\'e potentials and interlayer tunneling}

The continuum model~\eqref{eq:continuum} contains a quadratic intralayer kinetic energy centered at $\bm\kappa_{t,b}$ and smooth moir\'e-periodic intralayer potentials and tunneling. In momentum space the intralayer dispersion is
\begin{equation}\label{eq:ekin}
E_{\mathrm{kin},l}(\bm k)=\frac{1}{2m}|\bm k-\bm\kappa_l|^2~,
\qquad \bm\kappa_t\equiv\bm\kappa_{+}~,\ \ \bm\kappa_b\equiv\bm\kappa_{-}~,
\end{equation}
where the layer-dependent shifts $\bm\kappa_{\pm}$ arise from zone folding of the monolayer $K$ points under the moir\'e superlattice.

\paragraph*{First-harmonic truncation.} Because moir\'e fields are smooth on the scale of $a_M$, we retain only the six shortest reciprocal vectors $\{\mathbf{g}_i\}_{i=1}^6$ (with $\mathbf{g}_{i+3}=-\mathbf{g}_i$). The most general real intralayer potential at this order is
\begin{equation}\label{eq:V_general_firststar}
V_l(\bm r)= -\sum_{i=1}^{6} V_{\mathbf{g}_i,l}\,e^{i\phi_{\mathbf{g}_i,l}}\,e^{i\mathbf{g}_i\cdot \bm r}~,
\end{equation}
with $V_{\mathbf{g},l}=V_{-\mathbf{g},l}$ and $\phi_{-\mathbf{g},l}=-\phi_{\mathbf{g},l}$ following from reality.

\paragraph*{$C_{3z}$ constraint.} Under threefold rotation $\mathcal R_{2\pi/3}$ about the out-of-plane axis, the continuum model satisfies
$V_l(\mathcal R_{2\pi/3}\bm r)=V_l(\bm r)$.
At lowest harmonic order this enforces equality of the three independent Fourier amplitudes related by rotation,
\begin{equation}
V_{\mathbf{g}_1,l}=V_{\mathbf{g}_3,l}=V_{\mathbf{g}_5,l}\equiv V~,\qquad
\phi_{\mathbf{g}_1,l}=\phi_{\mathbf{g}_3,l}=\phi_{\mathbf{g}_5,l}\equiv \phi_l~,
\end{equation}
and similarly for their negatives by reality.
Using $\mathbf{g}_{1,3,5}$ for the three wavevectors separated by $120^\circ$ (and $\mathbf{g}_{2,4,6}=-\mathbf{g}_{1,3,5}$), Eq.~\eqref{eq:V_general_firststar} reduces to the standard cosine form
\begin{equation}\label{eq:V_cos}
V_l(\bm r)=-2V\sum_{i=1,3,5}\cos\big(\mathbf{g}_i\cdot \bm r+\phi_l\big)~.
\end{equation}
We fix the origin of $\bm r$ at the center of an MM-stacking region; with this convention $\phi_l$ parametrizes the relative registry of the moir\'e modulation within each layer.

\paragraph*{Twofold layer-exchange symmetry.} For twisted homobilayers, a twofold operation that maps $\bm r\mapsto-\bm r$ interchanges the layers. At the level of Eq.~\eqref{eq:continuum} this implies
\begin{equation}\label{eq:C2_constraint}
V_1(\bm r)=V_2(-\bm r)~, \qquad T(\bm r)=T^{\dagger}(-\bm r)~,
\end{equation}
up to layer-dependent overall phases. Applied to Eq.~\eqref{eq:V_cos}, this fixes equal amplitudes and opposite phases,
\begin{equation}\label{eq:phi_relation}
\phi_2=-\phi_1\equiv \phi~.
\end{equation}

\paragraph*{Interlayer tunneling.} The interlayer tunneling $T(\bm r)$ is also moir\'e-periodic and, to the same lowest-harmonic accuracy, is dominated by three momenta connecting the folded layer valleys. Imposing $C_{3z}$ symmetry fixes the three tunneling harmonics to have equal magnitude; the remaining overall complex phase can be absorbed by a redefinition of the layer spinors. In the gauge consistent with Eq.~\eqref{eq:continuum}, we therefore take
\begin{equation}\label{eq:Tmoire}
T(\bm r)=w\left(1+e^{i\mathbf{g}_2\cdot \bm r}+e^{i\mathbf{g}_3\cdot \bm r}\right)~,
\end{equation}
where $w\ge 0$ sets the tunneling scale. Different (but equivalent) conventions correspond to cyclic permutations of $\mathbf{g}_{1,2,3}$ and/or a shift of the real-space origin.

\paragraph*{Parameters and symmetry.} The continuum parameters $(m,V,w,\phi)$ can be extracted from first-principles fits and depend on material choice and screening. Spin-orbit coupling reduces spin symmetry to a conserved out-of-plane component; correspondingly, the single-valley Hamiltonian has a $U(1)$ spin symmetry generated by $S_z$ rather than full $SU(2)$, consistent with Ising-like ferromagnetism in many moir\'e TMD settings.

\section{Skyrmion Chern Band Model and Emergent Inhomogeneous Gauge Field}
\label{sec:skyrmion_chern_band}

To capture the essential topological content and to connect moir\'e Chern bands in twisted transition-metal dichalcogenide homobilayers to a Landau-level-like description while keeping moir\'e periodicity explicit, we employ a continuum ``skyrmion Chern band'' setting in which electrons couple to a smooth moir\'e-periodic pseudospin texture~\cite{PhysRevLett.132.096602,PhysRevB.110.035130,qxnw-8q4y}.  The topology arises from a slowly varying two-component (layer/spin) pseudospin texture and includes, as a special case, the adiabatic continuum description often used for twisted MoTe$_2$. 

At the single-particle level, adiabatic motion in the texture-aligned low-energy branch generates an emergent real-space Berry connection with quantized flux per moir\'e unit cell, yielding a Chern band in the adiabatic limit. In interacting moir\'e systems, such textures can arise self-consistently from interactions; here we take the texture, equivalently the local splitting field, as a given background/saddle and develop a controlled projection that defines the corresponding projected interacting theory.

\subsection{Continuum model and pseudospin texture}

The continuum model for a twisted TMD homobilayer, when restricted to a single valley/spin sector (e.g.\ $s =\uparrow$), we write the interacting Hamiltonian as
\begin{equation}
H = \int d^2r\;\psi^\dagger(\bm r)\,\mathcal H_{\uparrow}(\bm r)\,\psi(\bm r) + H_C~,
\end{equation}
where $H_C$ is the Coulomb interaction defined in Sec.~\ref{sec:continuum_model}. Up to layer-dependent plane-wave transformations that reshuffle the momentum offsets $\bm\kappa_\pm$ into phases of the off-diagonal tunneling, the single-particle Hamiltonian can be cast in the Pauli-matrix form
\begin{equation}\label{eq:H_continuum_local}
\mathcal H_{\uparrow}(\bm r)=\frac{\bm p^2}{2m}\sigma_0
- J(\bm r)\,\hat{\bm n}(\bm r)\cdot\bm\sigma
+ V(\bm r)\,\sigma_0~.
\end{equation}
Here $m$ is an effective mass, $(\sigma_0,\, \bm\sigma)$ are the identity and Pauli matrices in the layer-pseudospin space, $J(\bm r)\ge 0$ sets the local pseudospin exchange, $\hat{\bm n}(\bm r)$ is a smooth moir\'e-periodic unit vector field, and $V(\bm r)$ is a smooth scalar moir\'e potential with the periodicity of the moir\'e superlattice. A convenient identification consistent with Eq.~\eqref{eq:continuum} is
\begin{align}\label{eq:J_identification}
\begin{split}
&J_x(\bm r)-iJ_y(\bm r)\equiv - T(\bm r)~,\qquad
J_x(\bm r)+iJ_y(\bm r)\equiv - T^\dagger(\bm r)~,\\
&J_z(\bm r)\equiv \frac{V_2(\bm r)-V_1(\bm r)}{2}~,\qquad
V(\bm r)\equiv \frac{V_1(\bm r)+V_2(\bm r)}{2}~,    
\end{split}
\end{align}
up to overall constants and gauge conventions.

\paragraph*{Interpretation of $J(\bm r)\hat{\bm n}(\bm r)$.} Eq.~\eqref{eq:H_continuum_local} should be viewed as an effective one-body description in the presence of a slowly varying pseudospin splitting field $J(\bm r)\hat{\bm n}(\bm r)$, which locally separates two pseudospin branches by an energy $\sim 2J(\bm r)$. In microscopic moir\'e models its components can be identified with tunneling and layer-asymmetric potentials, Eq.~\eqref{eq:J_identification}, while in interaction-driven settings $J\hat{\bm n}$ may additionally represent a self-consistent pseudospin order parameter. Our controlled expansion below assumes only that $J$ is large compared to moir\'e kinetic scales and that the texture varies slowly; it does not rely on the microscopic mechanism that produces the texture.

We consider chiral textures $\hat{\bm n}(\bm r)$ forming a periodic skyrmion lattice
\begin{equation}
\hat{\bm n}(\bm r+\bm a_i)=\hat{\bm n}(\bm r)~,    
\end{equation}
with nonzero winding number (Pontryagin number) per moir\'e unit cell. In this regime, the lowest band can be understood as an adiabatic skyrmion Chern band: the pseudospin follows $\hat{\bm n}(\bm r)$ and the orbital motion experiences an emergent gauge field. Such emergent fields contain an integer number of flux quanta per unit cell. Analogous effective magnetic fields arising from non-collinear spin textures, which have been extensively explored in magnetic metals, are known to produce the topological Hall effect. However, studies of skyrmion textures in low-density semiconductors remain relatively sparse. In this semiconductor context, the topological Hall effect manifests distinctly as a quantized anomalous Hall effect resulting from the formation of topologically nontrivial Chern bands. 

\paragraph*{Gauge transformation and adiabatic projection.} Let $U(\bm r)\in SU(2)$ be a smooth local pseudospin rotation that aligns the texture with the
$z$ axis,
\begin{equation}
U^\dagger(\bm r)\,\hat{\bm n}(\bm r)\cdot\bm\sigma\,U(\bm r)=\sigma_z~.
\end{equation}
In the rotated frame, the gradients of $U(\bm r)$ generate a non-Abelian connection
\begin{equation}
\mathcal A_i(\bm r)=iU^\dagger(\bm r)\,\partial_i U(\bm r)~,
\end{equation}
so that the kinetic term acquires $\bm p\rightarrow \bm p-\bm{\mathcal A}(\bm r)$. In the $\sigma_z$ basis, the diagonal component
\begin{equation}
a_i(\bm r)\equiv (\mathcal A_i)_{\uparrow\uparrow}
=- i\langle u(\bm r)|\partial_i u(\bm r)\rangle~, \qquad |u(\bm r)\rangle = U(\bm r)|\uparrow\rangle~,
\end{equation}
acts as an emergent $U(1)$ gauge field within the adiabatic subspace, while the off-diagonal components mediate inter-branch mixing.

In the large-splitting regime, the low-energy wavefunctions are locally aligned with $\hat{\bm n}(\bm r)$. Projecting onto the corresponding branch yields an effective single-branch one-body Hamiltonian
\begin{equation}\label{eq:Heff_U1}
H_{1{\rm b},\rm eff}=\frac{\big[\bm p-\bm a(\bm r)\big]^2}{2m}+V(\bm r)+\Phi_{\rm g}(\bm r)+\cdots~,
\end{equation}
where $V(\bm r)$ is the smooth moir\'e scalar potential in Eq.~\eqref{eq:H_continuum_local} and
\begin{equation}
\Phi_{\rm g}(\bm r)=\frac{1}{8m}\big(\partial_i \hat{\bm n}\big)^2
\end{equation}
is the geometric scalar potential generated by virtual transitions to the opposite branch (the systematic $1/J$ correction is developed in Sec.~\ref{sec:controlled_expansions}).

The emergent magnetic field associated with $a_i$ is
\begin{equation}\label{eq:btex}
b(\bm r)=\varepsilon^{ij}\partial_i a_j(\bm r)
=\frac{1}{2}\hat{\bm n}(\bm r)\cdot\Big[\partial_x\hat{\bm n}(\bm r)\times\partial_y\hat{\bm n}(\bm r)\Big]~,
\end{equation}
whose flux through a moir\'e unit cell $A_M$ is quantized by the skyrmion number,
\begin{equation}\label{eq:flux_quant_polished}
\frac{1}{2\pi}\int_{A_M} d^2r\; b(\bm r)=Q\in\mathbb Z~.
\end{equation}
In the strict adiabatic limit, the polarized miniband inherits a Chern number $C=\pm Q$, with the sign set by the choice of branch. It is often convenient to decompose the emergent field into a uniform background plus moir\'e-periodic modulations,
\begin{equation}
b(\bm r)=b_0+\delta b(\bm r)~,\qquad b_0 \equiv \frac{1}{A_M}\int_{A_M} d^2r\; b(\bm r)=\frac{2\pi Q}{A_M}~,
\end{equation}
and treat $\delta b(\bm r)$ on the same footing as other periodic moir\'e scale perturbations.

A controlled organization of corrections beyond strict adiabaticity is obtained from a systematic Schrieffer-Wolff expansion in gradients and $1/J$. The gauge-covariant formulation, quantum-geometric identities, and the explicit SW reduction used below are developed in Sec.~\ref{sec:controlled_expansions}.

\paragraph*{Projected interacting theory.} When interactions $H_C$ are included, the low-energy many-body problem is obtained by expressing $H_C$ in terms of the projected (and, at finite $J$, dressed) density operator in the low-energy branch. The controlled SW construction in Sec.~\ref{sec:controlled_expansions} produces both the effective single-branch Hamiltonian and the corresponding operator dressing; together they define the interacting projected theory used in our response and EFT analyses. The remainder of this section focuses on the emergent $U(1)$ gauge structure and flux quantization that control the adiabatic-limit band topology.

\section{Controlled Expansions beyond Adiabaticity}
\label{sec:controlled_expansions}

\begin{figure*}[t]
  \centering
  \includegraphics[width=\linewidth]{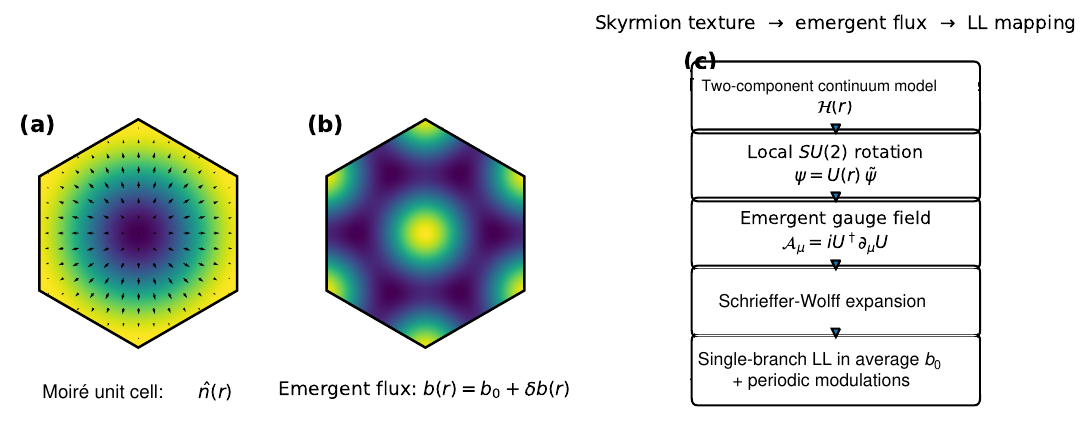}
  \caption{
  (a) Moir\'e unit cell with a skyrmion-like pseudospin texture $\hat{\bm n}(\bm r)$.
  (b) Emergent flux density $b(\bm r)$ illustrating the decomposition $b(\bm r)=b_0+\delta b(\bm r)$.
  (c) Controlled mapping: local $SU(2)$ rotation $\rightarrow$ emergent gauge field $\mathcal A_\mu = i U^\dagger \partial_\mu U$ $\rightarrow$ Schrieffer-Wolff expansion (controlled $1/J$) $\rightarrow$ single-branch Landau-level description in the average flux $b_0$ with periodic modulations.}
  \label{fig:1}
\end{figure*}

This section develops the controlled microscopic step underlying the rest of the paper: an operator-level Schrieffer-Wolff reduction in $1/J$ that integrates out the high-energy pseudospin branch and yields an effective single-branch Hamiltonian and systematically dressed physical operators, e.g. density, current, etc., acting within the low-energy branch, see Fig.~\ref{fig:1}. Together these results define the projected interacting low-energy theory once interactions are expressed in terms of the dressed density operator.

After deriving the $1/J$ expansion in a manifestly gauge-covariant form, we reorganize the single-branch problem as Landau levels in the spatially averaged emergent field $b_0$ perturbed by moir\'e-periodic magnetic and scalar modulations. This ``uniform-plus-periodic'' representation provides a practical starting point for magnetic-Bloch diagonalization and for response calculations that keep moir\'e Umklapp structure explicit. When we later compute collective spectra, we will employ a standard time-dependent Hartree-Fock/random-phase (TDHF/RPA) approximation in this basis; this approximation is conceptually distinct from the controlled $1/J$ projection developed here.

\subsection{Rotation to the local frame and emergent gauge fields}

We start from the texture Hamiltonian~\eqref{eq:H_continuum_local},
\begin{equation}
H=\frac{\bm p^2}{2m}\sigma_0 - J(\bm r)\,\hat{\bm n}(\bm r)\cdot \bm\sigma + V(\bm r)\,\sigma_0~,
\end{equation}
with $J(\bm r)\ge 0$ and smooth, moir\'e-periodic $\hat{\bm n}(\bm r)$ and $V(\bm r)$. (Relative sign conventions can be absorbed into $\hat{\bm n}\to -\hat{\bm n}$; we choose \eqref{eq:H_continuum_local} such that the locally aligned pseudospin is the low-energy branch.) We recall that $J(\bm r)\hat{\bm n}(\bm r)$ denotes a generic slowly varying branch-splitting field; see Sec.~\ref{sec:skyrmion_chern_band}.

Let $U(\bm r)\in SU(2)$ be a smooth rotation aligning the local exchange axis with $\sigma_z$,
\begin{equation}
U^\dagger(\bm r)\,\hat{\bm n}(\bm r)\cdot\bm\sigma \, U(\bm r)=\sigma_z~.
\end{equation}
In the rotated frame $\psi(\bm r)=U(\bm r)\tilde\psi(\bm r)$, we use
$U^\dagger \bm p\, U=\bm p-\bm{\mathcal A}(\bm r)$ with the emergent non-Abelian gauge field
\begin{equation}\label{eq:nonabelian_A}
\mathcal A_i(\bm r)= i U^\dagger(\bm r)\,\partial_i U(\bm r)
=\mathcal A_i^a(\bm r)\,\sigma_a \in \mathfrak{su}(2)~,
\end{equation}
to obtain
\begin{equation}\label{eq:H_rotated}
\tilde H = \frac{\big[\bm p-\bm{\mathcal A}(\bm r)\big]^2}{2m} - J(\bm r)\,\sigma_z +V(\bm r)\,\sigma_0~.
\end{equation}

\paragraph*{Residual $U(1)$ gauge freedom.} The choice of local frame is not unique: the rotation
\begin{equation}\label{eq:residual_U1}
U(\bm r)\ \to\ U(\bm r)\,e^{i\varphi(\bm r)\sigma_z/2}
\end{equation}
leaves $U^\dagger \hat{\bm n}\cdot\bm\sigma\,U=\sigma_z$ invariant. Under~\eqref{eq:residual_U1}, the diagonal Abelian component shifts by a gradient while the off-diagonal components carry unit $U(1)$ charge.

For later use we decompose $\mathcal A_i$ in the $\sigma_z$ basis,
\begin{equation}\label{eq:A_decomp}
\mathcal A_i(\bm r)=
\begin{pmatrix}
a_i(\bm r) & W_i(\bm r)\\
W_i^\dagger(\bm r) & -a_i(\bm r)
\end{pmatrix},
\qquad
a_i\equiv (\mathcal A_i)_{++}~,
\quad
W_i\equiv (\mathcal A_i)_{+-}~.
\end{equation}
Under~\eqref{eq:residual_U1},
\begin{equation}\label{eq:gauge_transforms_aW}
a_i(\bm r)\to a_i(\bm r)+\tfrac{1}{2}\partial_i\varphi(\bm r)~,
\qquad
W_i(\bm r)\to e^{-i\varphi(\bm r)}\,W_i(\bm r)~.
\end{equation}

Projecting onto the local $\sigma_z=\pm1$ branch produces an Abelian Berry connection $\pm a_i(\bm r)$ and emergent magnetic field
\begin{equation}\label{eq:Beff_pm}
b^{(\pm)}(\bm r)=\varepsilon^{ij}\partial_i\big(\pm a_j(\bm r)\big)
=\pm \frac{1}{2}\hat{\bm n}\cdot(\partial_x\hat{\bm n}\times\partial_y\hat{\bm n})~.
\end{equation}
For a skyrmion lattice, the flux of $b^{(\pm)}$ through a moir\'e unit cell is quantized to $\pm 2\pi Q$ with integer skyrmion number.

\subsection{Quantum geometric tensor}

The real-space quantum geometric tensor of the local spinor $|u_+(\bm r)\rangle$ is
\begin{equation}\label{eq:QGT_def}
\mathcal Q_{ij}(\bm r)\equiv
\langle \partial_i u_+(\bm r)|(1-|u_+(\bm r)\rangle\langle u_+(\bm r)|)|\partial_j u_+(\bm r)\rangle
\equiv g_{ij}(\bm r)+\frac{i}{2}\Omega_{ij}(\bm r)~.
\end{equation}
For a two-level system, $(1-|u_+\rangle\langle u_+|)=|u_-\rangle\langle u_-|$. Using $\<{u_-|\partial_j u_+}\>=-\<{\partial_j u_-|u_+}\>$ and the definition $W_i=(\mathcal A_i)_{+-}=i\<{u_+|\partial_i u_-}\>$, one finds the compact identity
\begin{equation}\label{eq:QGT_equals_WW}
\mathcal Q_{ij}(\bm r)=W_i(\bm r)\,W_j^\dagger(\bm r)~,
\end{equation}
and therefore the metric and Berry curvature are
\begin{equation}\label{eq:g_Omega_from_T}
g_{ij}(\bm r)= \Re \, \mathcal Q_{ij}(\bm r)~,
\qquad
\Omega_{ij}(\bm r)=2 \Im \, \mathcal Q_{ij}(\bm r)~.
\end{equation}

Parameterizing the texture by polar angles,
\begin{equation}\label{eq:n_theta_phi}
\hat{\bm n}(\bm r)=\big(\sin\theta(\bm r)\cos\phi(\bm r),\,\sin\theta(\bm r)\sin\phi(\bm r),\,\cos\theta(\bm r)\big)~,
\end{equation}
a standard choice for the local eigen-spinor is
\begin{equation}\label{eq:uplus_spinor}
|u_+(\bm r)\rangle=
\begin{pmatrix}
\cos\frac{\theta(\bm r)}{2}\\
e^{i\phi(\bm r)}\sin\frac{\theta(\bm r)}{2}
\end{pmatrix}~.
\end{equation}
This gives
\begin{align}
a_i(\bm r)
&= -i\langle u_+|\partial_i u_+\rangle
=\frac{1-\cos\theta(\bm r)}{2}\partial_i\phi(\bm r)~,
\label{eq:ai_theta_phi}\\
W_i(\bm r)
&= (\mathcal A_i)_{+-}
= -\frac{e^{-i\phi(\bm r)}}{2}\Big(\sin\theta(\bm r)\partial_i\phi(\bm r)+i\partial_i\theta(\bm r)\Big)~.
\label{eq:Wi_theta_phi}
\end{align}
From Eq.~\eqref{eq:QGT_equals_WW} one obtains the standard geometric identities
\begin{align}
g_{ij}(\bm r)
&=\frac{1}{4}\Big(\partial_i\theta\partial_j\theta+\sin^2\theta\partial_i\phi\partial_j\phi\Big)
=\frac{1}{4}\partial_i\hat{\bm n}\cdot\partial_j\hat{\bm n}~,
\label{eq:gij_n}\\
\Omega_{xy}(\bm r)
&=\frac{1}{2}\sin\theta\big(\partial_x\theta\partial_y\phi-\partial_y\theta\partial_x\phi\big)
=\frac{1}{2}\hat{\bm n}\cdot(\partial_x\hat{\bm n}\times\partial_y\hat{\bm n})~.
\label{eq:Omega_n}
\end{align}

\subsection{Schrieffer-Wolff transformation and controlled \texorpdfstring{$1/J$}{1/J} expansion}

The rotated-frame Hamiltonian~\eqref{eq:H_rotated} describes two pseudospin branches split by $2J(\bm r)$ and coupled by the off-diagonal components of the emergent non-Abelian gauge field $\mathcal A_i$. The strict adiabatic limit corresponds to $J\to\infty$, where the low-energy dynamics is confined to a single branch and the gauge structure reduces to the diagonal $U(1)$ connection $a_i$. At large but finite $J$, non-adiabatic effects refer to controlled virtual transitions into the high-energy branch induced by the off-diagonal gauge field $W_i$, and can be integrated out systematically in powers of $1/J$.

\paragraph*{Block structure.} Writing $\tilde H$ in the $\sigma_z$ eigenbasis,
\begin{equation}
\tilde H=
\begin{pmatrix}
H_{++} & H_{+-}\\
H_{-+} & H_{--}
\end{pmatrix}~,
\qquad H_{-+}=H_{+-}^\dagger~,
\end{equation}
one finds
\begin{align}
H_{++}
&=
\frac{1}{2m}\sum_i\Big[(\pi_i^{(+)})^2 + W_i W_i^\dagger\Big]
+V(\bm r)-J(\bm r)~,
\label{eq:Hpp_final}
\\
H_{--}
&=
\frac{1}{2m}\sum_i\Big[(\pi_i^{(-)})^2 + W_i^\dagger W_i\Big]
+V(\bm r)+J(\bm r)~,
\\
H_{+-}
&=
-\frac{1}{2m}\sum_i\Big(\pi_i^{(+)}W_i+W_i\pi_i^{(-)}\Big)~.
\label{eq:Hpm_final}
\end{align}
Here we introduced branch-covariant momenta
\begin{equation}\label{eq:branch_covariant_pi}
\pi_i^{(+)}\equiv -i\partial_i-a_i(\bm r)~,
\qquad
\pi_i^{(-)}\equiv -i\partial_i+a_i(\bm r)~,
\end{equation}
and the full covariant derivative is
\begin{equation}
D_i = -i\partial_i-\mathcal A_i=
\begin{pmatrix}
\pi_i^{(+)} & -W_i\\
- W_i^\dagger & \pi_i^{(-)}
\end{pmatrix}~,
\end{equation}
so that
\begin{align}
(\bm p-\bm{\mathcal A})^2
&=\sum_i D_iD_i \nonumber\\
&=\sum_i
\begin{pmatrix}
(\pi_i^{(+)})^2 + W_i W_i^\dagger
&
-\big(\pi_i^{(+)}W_i+W_i\pi_i^{(-)}\big)
\\[3pt]
-\big(\pi_i^{(-)}W_i^\dagger+W_i^\dagger\pi_i^{(+)}\big)
&
(\pi_i^{(-)})^2 + W_i^\dagger W_i
\end{pmatrix}~,
\label{eq:D2_blocks}
\end{align}
which is manifestly covariant under the residual $U(1)$ gauge freedom~\eqref{eq:residual_U1}.

The diagonal geometric scalar potential is
\begin{equation}\label{eq:BO_potential}
\Phi_{\rm g}(\bm r)\equiv \frac{1}{2m}\sum_i W_i(\bm r)W_i^\dagger(\bm r)
=\frac{1}{2m} \Tr g(\bm r)~,
\end{equation}
where $g_{ij}$ is the real-space quantum metric of the local adiabatic spinor, and reduces, for a smooth two-level texture, to $\Phi_{\rm g}=\frac{1}{8m}(\partial_i\hat{\bm n})^2$ up to conventions for which branch is the low-energy one.

\paragraph*{Controlled regime.} We assume a hierarchy in which the exchange splitting is the largest scale,
\begin{equation}
J \gg E_M~, \qquad J \gg \omega_c~, \qquad g\ell_{b_0}\ll 1~,
\end{equation}
where the moir\'e kinetic scale is $E_M\sim 1/(m  a_M^2)$ and $\omega_c=|b_0|/m$ is the cyclotron energy associated with the average emergent field
\begin{equation}
b_0 \equiv \frac{1}{A_M}\int_{A_M} d^2r\,b(\bm r)~, \qquad \ell_{b_0}^2 \equiv 1/|b_0|~,
\end{equation}
$\ell_{b_0}$ is the corresponding magnetic length, and $\mathbf g$ is a typical moir\'e reciprocal-lattice vector with $g\equiv |\mathbf g|$. The parameter $g\ell_{b_0} \ll 1$ controls the long-wavelength/gradient expansion of projected-operator algebra, e.g. Moyal-product deformations. In realistic moir\'e settings it can be $\mathcal{O}(1)$; we therefore use this expansion as an organizing principle for analytic formulas, while the Umklapp formulation does not rely on $g\ell_{b_0}\ll 1$.

\paragraph*{Schrieffer-Wolff reduction.} Because the branch splitting satisfies
\begin{equation}
H_{--}-H_{++}=2J(\bm r)+O(E_M)~,
\end{equation}
the low-energy ($+$) sector is separated from the high-energy ($-$) sector by a large gap: $\Delta(\bm r)\simeq 2J(\bm r)$. A resolvent form of the Schrieffer-Wolff reduction gives an effective Hamiltonian in the $+$ subspace,
\begin{equation}\label{eq:Heff_resolvent}
H_{\rm eff}(E)= H_{++} - H_{+-}\,\frac{1}{H_{--}-E}\,H_{-+}~.
\end{equation}
Expanding $\left(H_{--}-E\right)^{-1}$ in $1/J$ yields, to leading nontrivial order,
\begin{equation}\label{eq:HeffSWfinal}
H_{\rm eff}= H_{++} - H_{+-}\,\frac{1}{2J(\bm r)}\,H_{-+} +O(J^{-2})~.
\end{equation}
An equivalent expression follows from the standard SW-generator construction; for completeness we summarize it in Appendix~\ref{app:SW_generator}. 

When $J(\bm r)$ varies, $J(\bm r)$ does not commute with gradients inside $H_{-+}$; a manifestly Hermitian choice is to symmetrize the $1/J$ insertion,
\begin{equation}\label{eq:Heff_1overJ_sym}
H_{\rm eff}=H_{++} -\frac12\left\{\frac{1}{2J(\bm r)},\,H_{+-}H_{-+}\right\}+O(J^{-2})~,
\qquad \{A,B\}\equiv AB+BA~.
\end{equation}
This differs from~\eqref{eq:HeffSWfinal} by commutator/ordering terms that are higher order in gradients, $\sim [J^{-1},\, H_{+-}]H_{-+}$.

\begin{figure}[t]
  \centering
  \includegraphics[width=1\linewidth]{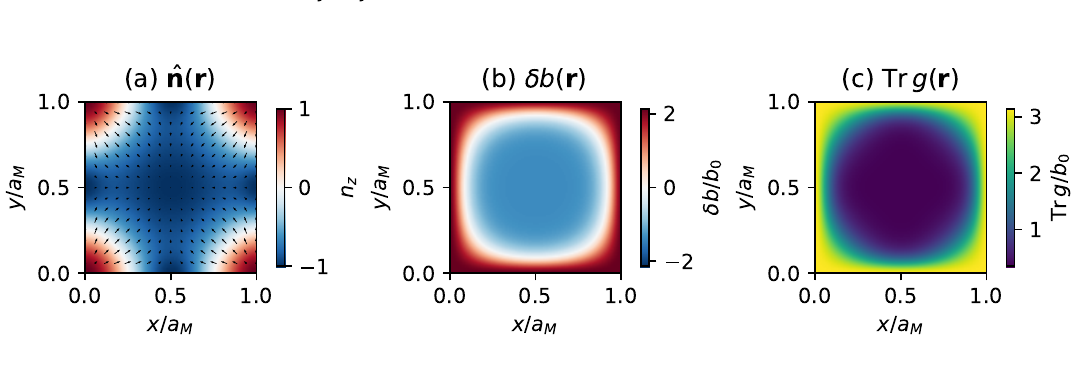}
  \caption{Real-space texture, emergent flux, and quantum geometry.
  (a) Example periodic skyrmion pseudospin texture $\hat{\bm n}(\bm r)$ (color: $n_z$, arrows: in-plane components).
  (b) Emergent flux density $b(\bm r)=\varepsilon^{ij}\partial_i a_j(\bm r)=\tfrac12\hat{\bm n}\cdot(\partial_x\hat{\bm n}\times\partial_y\hat{\bm n})$, decomposed as $b=b_0+\delta b$ with quantized unit-cell flux.
  (c) Trace of the real-space quantum metric $\mathrm{Tr}\,g(\bm r)=\frac14[(\partial_x\hat{\bm n})^2+(\partial_y\hat{\bm n})^2]$,
  illustrating the pointwise two-level inequality $\mathrm{Tr}\,g(\bm r)\ge |b(\bm r)|$ and the geometric origin of flux inhomogeneity.}
  \label{fig2}
\end{figure}

\paragraph*{Leading non-adiabatic correction and quantum geometry.} Substituting~\eqref{eq:Hpm_final} into~\eqref{eq:HeffSWfinal} yields the leading correction from virtual branch mixing,
\begin{widetext}
\begin{align}
\delta H_{1/J}
&\equiv
-H_{+-}\,\frac{1}{2J(\bm r)}\,H_{-+}
\nonumber\\
&=-\frac{1}{8 m^2}\sum_{i,j}\Big(\pi_i^{(+)}W_i+W_i\pi_i^{(-)}\Big)\,\frac{1}{J(\bm r)}\,\Big(\pi_j^{(-)}W_j^\dagger+W_j^\dagger\pi_j^{(+)}\Big) +O(J^{-2})~.
\label{eq:deltaH_1overJ_exact}
\end{align}
\end{widetext}
Using the quantum geometric tensor identity $\mathcal Q_{ij}(\bm r)\equiv W_iW_j^\dagger=g_{ij}+\frac{i}{2}\Omega_{ij}$ for the real-space quantum geometric tensor, and working to leading order in the long-wavelength/slow-modulation limit so that commutators with $1/J$ and higher derivatives are parametrically small, one can write the compact geometric form
\begin{equation}\label{eq:deltaH_1overJ}
\delta H_{1/J} \simeq -\frac{1}{2m^2} \pi_i^{(+)}\frac{\mathcal Q_{ij}(\bm r)}{J(\bm r)}\pi_j^{(+)} \simeq -\frac{1}{2m^2J(\bm r)}\,
\pi_i^{(+)}\Big[g_{ij}(\bm r)+\frac{i}{2}\Omega_{ij}(\bm r)\Big]\pi_j^{(+)} + \cdots~,
\end{equation}
which makes the physical content transparent:
finite $J$ induces controlled non-adiabatic corrections governed entirely by the local real-space quantum geometry of the texture, see Fig.~\ref{fig2}. In particular, the metric part renormalizes the parity-even kinetic structure in the projected branch, while the Berry-curvature part enters through the covariant-momentum algebra and produces additional gauge-covariant scalar/gradient contributions at the same order.

\paragraph*{Operator dressing and physical observables.} The Schrieffer-Wolff transformation that block-diagonalizes the Hamiltonian also dresses physical operators. Working in the rotated frame~\eqref{eq:H_rotated}, we split
\begin{equation}
\tilde H = H_0 + V~,\qquad H_0\equiv -J(\bm r)\,\sigma_z~,
\end{equation}
and decompose $V=V_{\rm d}+V_{\rm od}$ into parts that are diagonal/off-diagonal in the $\sigma_z$ basis.
The SW unitary transformation $e^{S}$ is generated by an anti-Hermitian, block-off-diagonal operator
\begin{equation}
S^\dagger=-S~,\qquad S=
\begin{pmatrix}
0 & S_{+-}\\
- S_{+-}^\dagger & 0
\end{pmatrix}~,
\end{equation}
chosen such that inter-branch mixing is removed order-by-order in $1/J$.
At leading order, $S$ is fixed by
\begin{equation}
[H_0,S]= V_{\rm od}~,
\end{equation}
which gives
\begin{equation}\label{eq:S_leading}
S_{+-} =- \frac{1}{2J(\bm r)}H_{+-} + O(J^{-2})~,
\end{equation}
with $H_{+-}$ given in~\eqref{eq:Hpm_final}. The low-energy branch is selected by the projector
\begin{equation}
P_+\equiv \frac{1+\sigma_z}{2}~,
\end{equation}
so that the effective Hamiltonian is $H_{\rm eff}=P_+ \, e^{S}\, \tilde H \, e^{-S} \, P_+$, reproducing~\eqref{eq:HeffSWfinal}.

Crucially, the same SW unitary must be applied to all microscopic operators. For any operator $\mathcal O$, e.g. density, current, stress, etc., the effective operator acting within the low-energy branch is
\begin{equation}\label{eq:Oeff_SW}
\mathcal O_{\rm eff}\equiv P_+\,e^{S}\,\mathcal O\,e^{-S}\,P_+~.
\end{equation}
Expanding the Baker-Campbell-Hausdorff series gives
\begin{equation}\label{eq:Oeff_BCH}
\mathcal O_{\rm eff} = P_+\,\mathcal O\,P_+ +P_+\,[S,\,\mathcal O]\,P_+
+\frac12 P_+\,[S,\,[S,\,\mathcal O]]\,P_+ +\cdots~,
\end{equation}
and therefore, to leading non-adiabatic order,
\begin{equation}\label{eq:Oeff_1overJ}
\mathcal O_{\rm eff} = P_+\,\mathcal O\,P_+ +P_+\,[S,\,\mathcal O]\, P_+ +O(J^{-2})~.
\end{equation}
The commutator term in~\eqref{eq:Oeff_1overJ} encodes virtual excursions into the high-energy pseudospin branch and is required for consistent, gauge-invariant physical response, e.g.\ charge/current operators entering Kubo formulas and optical sum rules.

\paragraph*{Projected interactions (many-body theory after SW).} For density-density interactions,
\begin{equation}
H_C=\frac{1}{2}\sum_{\bm q}V_C(\bm q)\,\rho_{-\bm q}\,\rho_{\bm q}~,
\end{equation}
the corresponding low-energy interacting Hamiltonian is obtained by applying the same SW unitary to the density operator,
\begin{equation}\label{eq:rho_eff_SW}
H_{C,{\rm eff}}=\frac{1}{2}\sum_{\bm q}V_C(\bm q)\,\rho^{\rm eff}_{-\bm q}\,\rho^{\rm eff}_{\bm q}~,\qquad \rho^{\rm eff}_{\bm q}\equiv P_+\,e^{S}\,\rho_{\bm q}\,e^{-S}\,P_+~.
\end{equation}
Thus the controlled output of the SW procedure is a well-defined projected interacting theory: interactions retain their standard form but act through systematically dressed density operators and form factors in the low-energy branch.

\paragraph*{Non-Abelian gauge field viewpoint.} Before projection, the local-frame rotation generates an $\mathfrak{su}(2)$ connection
$\mathcal A_\mu=iU^\dagger\,\partial_\mu U$.
Because it is induced by a change of basis, $\mathcal A_\mu$ is a pure gauge in the full two-component Hilbert space and its non-Abelian curvature vanishes identically,
$\mathcal F_{\mu\nu}\equiv \partial_\mu\mathcal A_\nu-\partial_\nu\mathcal A_\mu-i[\mathcal A_\mu,\mathcal A_\nu]=0$.
Nevertheless, the diagonal $U(1)$ component $a_\mu$ has a nontrivial field strength after projection, and the cancellation $\mathcal F_{\mu\nu}=0$ implies exact identities relating the projected emergent flux and quantum geometry to the off-diagonal gauge field $W_\mu$ that mediates branch mixing. For completeness we collect these non-Abelian identities and their relation to the projected $U(1)$ response in Appendix~\ref{app:nonabelian}.

\subsection{Uniform-plus-periodic decomposition and Landau-level representation}
\label{subsec:LL_representation}

For a skyrmion lattice with moir\'e unit-cell area $A_M$, flux quantization implies
\begin{equation}
\frac{1}{2\pi}\int_{A_M} d^2r\, b^{(+)}(\bm r)=C\in \mathbb Z~.
\end{equation}
We decompose the emergent field and connection into a uniform part and a periodic modulation,
\begin{equation}
b^{(+)}(\bm r)=b_0+\delta b(\bm r)~,\qquad
\int_{A_M} d^2r\,\delta b(\bm r)=0~,\qquad
b_0A_M=2\pi C~,
\end{equation}
and choose $a_i(\bm r)=a_{0,i}(\bm r)+\delta a_i(\bm r)$ with $\varepsilon^{ij}\partial_i a_{0,j}=b_0$ and $\varepsilon^{ij}\partial_i\delta a_j=\delta b$.

Expanding the periodic parts in moir\'e reciprocal vectors $\mathbf{g}\neq 0$,
\begin{equation}
\delta b(\bm r)=\sum_{\mathbf{g}\neq 0}\beta(\mathbf{g})\, e^{i\mathbf{g}\cdot\bm r}~, \qquad \delta a_i(\bm r)=\sum_{\mathbf{g}\neq 0}\alpha_i(\mathbf{g})\, e^{i\mathbf{g}\cdot\bm r}~,
\end{equation}
the Coulomb gauge $\nabla\cdot\delta\bm a=0$ fixes
\begin{equation}
\bm\alpha(\mathbf{g})= -i\frac{\hat{\bm z}\times \mathbf{g}}{|\mathbf{g}|^2}\beta(\mathbf{g})~, \qquad
\alpha_{\pm}(\mathbf{g})\equiv \alpha_x(\mathbf{g})\pm i\alpha_y(\mathbf{g}) =
\begin{cases}
\ \dfrac{g_x+ig_y}{|\mathbf{g}|^2}\,\beta(\mathbf{g})~, & (+)\\[6pt]
-\dfrac{g_x-ig_y}{|\mathbf{g}|^2}\,\beta(\mathbf{g})~, & (-)
\end{cases}
\label{eq:alpha_from_beta}
\end{equation}
We similarly Fourier-expand the scalar terms, including $V(\bm r)$ and $\Phi_{\rm g}(\bm r)$.

The projected single-branch Hamiltonian therefore describes a charged particle in a uniform field $b_0$ perturbed by a weak periodic modulation:
\begin{equation}
H= \frac{\big[\bm \Pi-\delta\bm a(\bm r)\big]^2}{2m} +V(\bm r)+\Phi_{\rm g}(\bm r)~, \qquad \bm \Pi\equiv \bm p-\bm a_0(\bm r)~.
\end{equation}
We introduce Landau-level ladder operators $a,a^\dagger$ associated with $b_0$ and magnetic length $\ell_{b_0}^2=1/|b_0|$, with cyclotron frequency $\omega_c=|b_0|/m$.
Using $\delta a_{\pm}(\bm r)=\delta a_x(\bm r)\pm i\delta a_y(\bm r)$ and Eq.~\eqref{eq:alpha_from_beta}, the Hamiltonian admits the operator expansion
\begin{widetext}
\begin{align}
H
&=\omega_c\left(a^\dagger a+\frac{1}{2}\right)-\frac{1}{\sqrt{2}\, m \ell_{b_0}}\sum_{\mathbf{g}\neq 0}\Big(a\,\alpha_+(\mathbf{g})+a^\dagger\,\alpha_-(\mathbf{g})\Big)\,e^{i\mathbf{g}\cdot\bm r}
\nonumber\\
&\quad
+\frac{1}{2m}\sum_{\mathbf{g},\mathbf{g}'}
\alpha_+(\mathbf{g})\alpha_-(\mathbf{g}')\,e^{i(\mathbf{g}+\mathbf{g}')\cdot\bm r}
+\sum_{\mathbf{g}}\big[V(\mathbf{g})+\Phi_{\rm g}(\mathbf{g})\big]\,e^{i\mathbf{g}\cdot\bm r}~.
\label{eq:H_LL_expansion}
\end{align}
\end{widetext}
This provides a practical starting point for magnetic Bloch diagonalization and for systematically retaining moir\'e Umklapp mixing in response calculations.

\subsection{Applicability beyond twisted TMD homobilayers} 

Although we motivate our analysis using a two-component layer-pseudospin continuum model appropriate to twisted TMD homobilayers, the central construction, a local unitary rotation into a texture-aligned frame followed by a controlled Schrieffer-Wolff projection in the regime of a large splitting $J$, only assumes a well-isolated low-energy branch (or manifold) and slow spatial variation of its internal eigenstates. For an $N$-component internal manifold (multiple layers, valleys, or orbitals), the same steps proceed with a local $U(N)$ transformation $U(\bm r)$ into the instantaneous eigenbasis, generating a non-Abelian gauge connection $\mathcal A_i(\bm r)=iU^\dagger\partial_iU$. Projecting onto a single isolated branch with local eigenvector $|u(\bm r)\rangle$ yields an Abelian Berry gauge connection $a_i(\bm r)=-i\langle u|\partial_i u\rangle$ and an emergent flux $b=\varepsilon^{ij}\partial_i a_j$, while the controlled beyond-adiabatic corrections generated by SW are governed by the corresponding real-space quantum geometric tensor $\mathcal Q_{ij}(\bm r)=\langle \partial_i u|(1-|u\rangle\langle u|)|\partial_j u\rangle$. When several nearly-degenerate branches are retained, the projected Berry connection and quantum geometry become genuinely matrix-valued (non-Abelian) within that subspace.

\paragraph*{Example: rhombohedral graphene aligned with hBN.} Recent work on rhombohedral $N$-layer graphene aligned with hBN (RNG/hBN) has identified analytically tractable ``ideal-limit'' regimes in which interactions can stabilize a periodic layer-pseudospin texture that can be viewed as a skyrmion lattice in $\mathbb{CP}^{N-1}$ parametrization and generates an emergent periodic magnetic field, yielding Landau-level-like Chern-band physics. Although RNG/hBN involves $N>2$ graphene layers, its low-energy sector is often effectively low-dimensional (in standard chiral descriptions, a two-band model with weight on outer-layer sublattices), while the ideal-limit formulation can be viewed as an isolated band endowed with a slowly varying internal layer spinor. In this language the texture is encoded by a normalized $N$-component layer spinor $\hat\chi(\bm r)$ (defined up to a local phase) and the associated rank-one projector $P(\bm r)=\hat\chi(\bm r)\hat\chi^\dagger(\bm r)$; the topological density may be written as $\frac{1}{2\pi i}\mathrm{Tr}\,P[\partial_xP,\partial_yP]$ and reduces to the usual Pontryagin density for $N=2$. From the standpoint of the present work, the essential ingredient is therefore not the literal number of graphene layers, but the existence of an isolated low-energy branch whose local internal state defines a texture and an associated Berry connection. Our controlled $1/J$ expansion, quantum-geometry diagnostics, and magneto-elastic effective theory can be adapted to such settings once the relevant gap scale and texture profile are specified. For $N=2$ the above reduces to the $SU(2)$ texture problem used throughout this work, with $P=(1+\hat{\bm n}\cdot\bm\sigma)/2$ and $b=\frac12\hat{\bm n}\cdot(\partial_x\hat{\bm n}\times\partial_y\hat{\bm n})$.

\section{Noncommutative Guiding-Center Kinematics from Skyrmion Flux}
\label{sec:noncommutative}

We now make an additional long-wavelength approximation by projecting the effective single-branch theory onto the LLL associated with the spatially averaged emergent field $b_0$. This LLL projection is controlled when the cyclotron gap set by $b_0$ is large compared to the retained moir\'e-periodic modulations and any Landau-level-mixing scales generated by interactions and finite-$J$ corrections. Throughout we treat $\delta b(\bm r)\equiv b(\bm r)-b_0$, scalar potentials, and the leading $1/J$ terms perturbatively in a gradient/harmonic expansion controlled by $|\mathbf g|\ell_{b_0}\ll 1$.

\paragraph*{Guiding-center decomposition.} For the uniform reference problem, decompose the position operator as
\begin{equation}
\bm r=\bm R+\bar{\bm R}~,
\end{equation}
where $\bm R$ is the guiding-center coordinate and $\bar{\bm R}$ is the cyclotron coordinate. Within the LLL only $\bm R$ remains dynamical, obeying the noncommutative algebra
\begin{equation}\label{eq:GC_comm_uniform}
[R_x,R_y]=-i\ell_{b_0}^2~, \qquad \ell_{b_0}^2\equiv \frac{1}{|b_0|}~,
\end{equation}
which is equivalent to the projective magnetic translation algebra and underlies the $W_\infty$/GMP structure of LLL-projected operators.

\paragraph*{Slowly varying flux and local noncommutativity.} In the skyrmion problem $b(\bm r)=b_0+\delta b(\bm r)$ is periodic. In the slow-modulation regime $g\ell_{b_0}\ll 1$ (with $g$ a typical moir\'e reciprocal vector) and for weak modulation $|\delta b|/b_0\ll 1$, the guiding-center symplectic form is smoothly deformed. To leading order in gradients one may parameterize this by a position-dependent noncommutativity parameter
\begin{equation}\label{eq:theta_of_R}
[R_x,R_y]= -i \theta(\bm R) + \mathcal O(\nabla^2)~,\qquad
\theta(\bm R) \equiv \frac{1}{b(\bm R)} =\ell_{b_0}^2\left[1-\frac{\delta b(\bm R)}{b_0}+\mathcal O(\delta b^2)\right]~.
\end{equation}
Higher-gradient corrections are organized in powers of $g\ell_{b_0}$ and can be included systematically.

\paragraph*{Star-product for LLL-projected operators.} For any smooth function $f(\bm r)$ define its LLL-projected operator
\begin{equation}
\overline{f}\equiv P_{\rm LLL}\,f(\bm r)\,P_{\rm LLL}~.
\end{equation}
In the long-wavelength limit, products of projected operators close under a deformation-quantization Moyal product,
\begin{equation}\label{eq:fg_star}
\overline{f}\,\overline{g}=\overline{f\star g}+\mathcal O(\nabla^3)~,
\end{equation}
where the underlying Poisson tensor is $\theta(\bm r)\varepsilon^{ij}$. To second order in $\theta$, including the leading $\nabla\theta$ correction, one may use
\begin{align}\label{eq:kontsevich_2D}
f\star g
&= f g +\frac{i}{2}\theta(\bm r)\varepsilon^{ij} \partial_i f \partial_j g
-\frac{1}{8}\theta(\bm r)^2\varepsilon^{ij}\varepsilon^{kl}\partial_i\partial_k f\partial_j\partial_l g
\nonumber\\
&\quad
-\frac{1}{12}\theta(\bm r)\partial_j\theta(\bm r)\varepsilon^{ij}\varepsilon^{kl}\Big[\partial_i\partial_k f\partial_l g-\partial_k f\partial_i\partial_l g\Big] +\cdots~.
\end{align}
This provides a practical analytic tool: once $b(\bm r)$ is specified by a few Fourier harmonics, the projected algebra can be computed order-by-order in $g\ell_{b_0}$.

\paragraph*{Physical versus guiding-center density.} The physical charge density entering response functions is the dressed density operator $\rho^{\rm eff}_{\bm q}$ obtained from the SW reduction, Eq.~\eqref{eq:rho_eff_SW}. When we further project to the LLL of $b_0$, we define the guiding-center density as $\bar\rho_{\bm q}\equiv P_{\rm LLL}\,\rho^{\rm eff}_{\bm q}\,P_{\rm LLL}$, which separates universal guiding-center kinematics from cyclotron form factors.

\paragraph*{Deformed $W_\infty$/GMP algebra.} It is convenient to separate the universal guiding-center algebra from cyclotron form factors. Define the guiding-center density operator
\begin{equation}
\bar\rho_{\bm q}\equiv e^{i\bm q\cdot\bm R}~.
\end{equation}
In the uniform limit $\theta(\bm r)\to \ell_{b_0}^2$, one has the exact product
\begin{equation}
\bar\rho_{\bm q}\,\bar\rho_{\bm k} =\exp\left[\frac{i\ell_{b_0}^2}{2} \bm q\wedge \bm k\right]\bar\rho_{\bm q+\bm k}~,
\end{equation}
and hence
\begin{equation}\label{eq:GMP_uniform_GC}
[\bar\rho_{\bm q},\bar\rho_{\bm k}] = 2i\sin\left(\frac{\ell_{b_0}^2}{2}\bm q\wedge \bm k\right)\bar\rho_{\bm q+\bm k}~.
\end{equation}
In the presence of a periodic emergent field, the Moyal-commutator yields a controlled deformation of the Girvin-Macdonald-Platzman algebra,
\begin{equation}\label{eq:GMP_deformed_def}
[\bar\rho_{\bm q},\bar\rho_{\bm k}] \simeq \left(e^{i\bm q\cdot\bm r}\star e^{i\bm k\cdot\bm r} - e^{i\bm k\cdot\bm r}\star e^{i\bm q\cdot\bm r}\right)_{\bm r\to \bm R}~.
\end{equation}
This deformation concerns the algebra of LLL-projected density operators at long wavelength; the many-body consequences depend on the interacting ground state realized within the projected theory. Expanding in $g\ell_{b_0}$, the leading term is a local GMP bracket controlled by the skyrmion flux density through $\theta(\bm R)=1/b(\bm R)$,
\begin{equation}\label{eq:GMP_local}
[\bar\rho_{\bm q},\bar\rho_{\bm k}] = 2i\sin\left(\frac{\theta(\bm R)}{2}\bm q\wedge \bm k\right)e^{i(\bm q+\bm k)\cdot\bm R} +\mathcal O(\nabla\theta)~.
\end{equation}
which is the central algebraic statement: the skyrmion flux texture induces a spatially modulated noncommutativity and hence a periodic deformation of the W$_\infty$/GMP algebra.

\paragraph*{Dipole kinematics of neutral excitations.} The noncommutative guiding-center algebra implies that density operators act as translations. For uniform $b_0$,
\begin{equation}
e^{i\bm q\cdot\bm R}\,\bm R\,e^{-i\bm q\cdot\bm R} = \bm R-\ell_{b_0}^2 \hat{\bm z}\times\bm q~,
\end{equation}
so a neutral excitation of momentum $\bm q$ carries a dipole moment $\bm d(\bm q)\propto -\ell_{b_0}^2\hat{\bm z}\times\bm q$. In the skyrmion case, Eq.~\eqref{eq:theta_of_R} yields the local generalization
\begin{equation}\label{eq:dipole_local}
e^{i\bm q\cdot\bm R}\,\bm R\,e^{-i\bm q\cdot\bm R} =\bm R-\theta(\bm R)\hat{\bm z}\times\bm q+\mathcal O(\nabla\theta)~.
\end{equation}
showing that flux/Berry-curvature inhomogeneity modulates dipolar kinematics.

\paragraph*{Non-adiabatic feature with controlled $1/J$ correction.} Beyond strict adiabaticity, the Schrieffer-Wolff expansion generates $O(J^{-1})$ terms that are explicitly controlled by the local quantum geometry of the texture (metric and Berry curvature). In the Landau-level language these terms act as a quantum-geometry-dependent Landau-level-mixing perturbation, and therefore renormalize the effective periodic potentials and generate additional higher-derivative corrections to the Moyal density algebra at fixed $b_0$. This provides a systematic route to quantify non-adiabatic deviations from ideal LLL-like kinematics in skyrmion Chern bands.

\subsection{Topological quantum-weight geometric bound and microscopic coefficient matching}

A useful diagnostic of how closely a Chern band realizes ideal Landau-level geometry is the small-momentum behavior of the physical equal-time charge structure factor,
\begin{equation}\label{eq:Sq_smallq}
S(\bm q) \equiv\frac{1}{A}\,\big\langle \rho_{\bm q}\, \rho_{-\bm q}\big\rangle
=\frac{1}{2\pi}\mathcal K_{ij}\,q_i q_j+\mathcal O(q^4)~,
\end{equation}
which defines the symmetric quantum-weight tensor $\mathcal K_{ij}$. Here $\langle\cdots\rangle$ denotes the ground-state expectation value in the interacting projected theory defined after SW (and before any further LLL idealization unless stated), and $\rho_{\bm q}$ is the corresponding dressed charge operator. In particular, $S(\bm q)$ refers to the physical charge correlator entering the optical sum rule, and should be distinguished from the guiding-center structure factor $\bar S(\bm q)=A^{-1}\langle \bar\rho_{\bm q}\,\bar\rho_{-\bm q}\rangle$ built from $\bar\rho_{\bm q}\equiv P_{\rm LLL}\, \rho_{\bm q}\, P_{\rm LLL}$.

The tensor $\mathcal K_{ij}$ is equivalently fixed by an optical sum rule,
\begin{equation}\label{eq:Kij_sumrule}
\mathcal K_{ij} = 2 \int_{0}^{\infty} d\omega\,\frac{\mathrm{Re}\,\sigma_{ij}(\omega)}{\omega}~,
\end{equation}
so $\mathcal K_{ij}$ measures the integrated longitudinal optical spectral weight at long wavelength. A recent general result~\cite{PhysRevX.14.011052} establishes a topological lower bound: for any gapped two-dimensional phase with Hall conductance $\sigma_{xy}$,
\begin{equation}\label{eq:K_bound}
\mathrm{Tr} \, \mathcal K \equiv \mathcal K_{xx}+\mathcal K_{yy} \ \ge\ |\kappa|~,
\qquad \kappa \equiv \frac{2\pi}{e^2} \sigma_{xy}~,
\end{equation}
where $\kappa$ is the dimensionless Hall response in units $e^2/h=1/2\pi$.
For an integer Chern insulator obtained by filling a band of Chern number $C$ one has $\kappa=C$, whereas for an Abelian fractional Hall state $\kappa$ is the corresponding rational Hall response, e.g., $\kappa=\nu$ for a Laughlin-type FCI in a $C=1$ band. In an isotropic system Eq.~\eqref{eq:K_bound} implies the structure-factor bound
\begin{equation}\label{eq:Sq_bound}
S(q)\ \ge\ \frac{|\kappa|}{4\pi}\,q^2~, \qquad q\to 0~.
\end{equation}
which provide an experimentally meaningful criterion for ``Landau-levelness'': the closer $\mathrm{Tr}\,\mathcal K$ is to $|\kappa|$, the closer the system is to saturating the topological minimum longitudinal spectral weight.

\paragraph*{Adiabatic skyrmion limit and near-saturation.} In the strict adiabatic skyrmion mapping, the low-energy single-branch problem reduces to a Landau-level system in a uniform emergent flux density $b_0$ plus parametrically weak periodic terms, with the single-particle Chern number fixed by flux quantization,
\begin{equation}\label{eq:C_from_b0}
C =\frac{1}{2\pi}\int_{A_M} d^2r\, b(\bm r) =\frac{b_0 A_M}{2\pi}\in\mathbb Z~,
\end{equation}
where $A_M$ is the moir\'e unit-cell area and $b(\bm r)=\varepsilon^{ij}\partial_i a_j$ is the emergent flux in the adiabatic subspace. In a continuum Landau-level setting (no periodic potential modulation), Kohn's theorem implies that the long-wavelength longitudinal optical weight is exhausted by the cyclotron mode, which leads to saturation of the bound. In our skyrmion setting, Eq.~\eqref{eq:C_from_b0} makes explicit why the adiabatic limit is the natural ``ideal geometry'' point: the emergent band approaches the Landau-level regime in which $\mathrm{Tr}\, \mathcal K$ is expected to be close to its topological minimum.

\paragraph*{Controlled sources of excess quantum weight.} Away from the ideal Landau-level limit, our microscopic formulation identifies two systematic mechanisms that generically contribute to the excess quantum weight $\mathrm{Tr}\,\mathcal K-|\kappa|$. First, flux inhomogeneity $b(\bm r)=b_0+\delta b(\bm r)$ deforms the guiding-center kinematics and, in a magnetic Bloch description, activates additional Umklapp-allowed dipole transitions. Second, finite exchange $J$ produces controlled non-adiabatic mixing through the Schrieffer-Wolff correction $\delta H_{1/J}$, i.e.\ branch mixing in the emergent-field language. Both effects therefore generically increase the integrated longitudinal optical weight entering Eq.~\eqref{eq:Kij_sumrule}. Accordingly, we organize the departure from ideal geometry as
\begin{equation}\label{eq:K_decomposition}
\mathrm{Tr}\,\mathcal{K} - |\kappa| \equiv \delta \mathcal K_{\delta b}
+\delta \mathcal K_{1/J} +\cdots \ge 0~,
\end{equation}
where the ellipsis includes higher-order and mixed contributions.

\paragraph*{A complementary real-space geometric bound from the texture.} The skyrmion mapping also yields a real-space geometric constraint on the geometric kinetic scalar potential,
\begin{equation}
\Phi_{\rm g}(\bm r)=\frac{1}{8m}(\partial_i\hat{\bm n})^2=\frac{1}{2m}\mathrm{Tr} \, g(\bm r)~,
\end{equation}
where $g_{ij}(\bm r)$ is the real-space quantum metric of the local adiabatic spinor. Using the local inequality $\mathrm{Tr}\, g(\bm r)\ge |\Omega_{xy}(\bm r)|$ together with $\int_{A_M}\Omega_{xy}=2\pi C$, one obtains the topology-controlled bound
\begin{equation}\label{eq:PhiBO_bound}
\overline{\Phi_{\rm g}} \equiv \frac{1}{A_M}\int_{A_M} d^2r\,\Phi_{\rm g}(\bm r) \ \ge\ \frac{\pi}{m} \frac{|C|}{A_M}~,
\end{equation}
which is a direct, texture-level constraint on a microscopic energetic term
generated by adiabatic projection, and it complements the momentum-space quantum-weight bound~\eqref{eq:K_bound}--\eqref{eq:Sq_bound}: the former constrains the minimal average scalar potential cost implied by nontrivial real-space geometry, while the latter constrains the minimal integrated longitudinal optical weight implied by the Hall topology. Together they provide a practical set of microscopic benchmarks for quantifying departures from the ideal Landau-level limit in skyrmion Chern band settings.

\section{Magnetic Bloch Basis and Density Form Factors}
\label{sec:magnetic_bloch}

We now formulate Umklapp-resolved density operators and response functions for the projected interacting theory defined after the controlled Schrieffer-Wolff reduction. The magnetic Bloch basis introduced below diagonalizes the effective one-body Hamiltonian in the low-energy branch (including moir\'e-periodic modulations), while interactions enter through the corresponding projected and dressed density operator. Collective-mode dispersions and optical spectral weights are then obtained within a standard matrix TDHF/RPA approximation; this approximation is conceptually distinct from the controlled $1/J$ projection and is used here as a practical scheme to capture Umklapp mixing.

We diagonalize the effective single-branch Hamiltonian in a magnetic Bloch basis associated with the uniform component $b_0$ of the emergent field and treat the periodic modulations from $\delta b(\bm r)$, scalar potentials, and $O(J^{-1})$ corrections within a controlled harmonic truncation. Let $\mathcal G$ denote the moir\'e reciprocal lattice. We parameterize a momentum transfer as
\begin{equation}\label{eq:Q_decomp}
\bm Q=\bm q+\bm G~, \qquad \bm q\in{\rm MBZ}~,\ \bm G\in\mathcal G~,
\end{equation}
where ${\rm MBZ}$ is the magnetic Brillouin zone and $\bm G$ labels moir\'e Umklapp sectors. 

Let $|\alpha,\bm k\rangle$ denote magnetic Bloch eigenstates, with $\bm k\in{\rm MBZ}$ and composite band index $\alpha$, including magnetic-sublattice and any Landau-level indices retained in a truncation. The Umklapp-resolved (dressed) density operator reads
\begin{equation}\label{eq:density_FF}
\rho_{\bm q+\bm G} \equiv \rho^{\rm eff}_{\bm q+\bm G}
= \sum_{\bm k\in{\rm MBZ}}\sum_{\alpha,\beta}
F^{\alpha\beta}_{\bm G}(\bm k;\bm q)\;
c^\dagger_{\alpha,\bm k}\,c_{\beta,\bm k+\bm q}~,
\end{equation}
where $\bm k+\bm q$ is understood modulo magnetic reciprocal vectors and folded back into the MBZ with density form factors
\begin{equation}\label{eq:FF_def}
F^{\alpha\beta}_{\bm G}(\bm k;\bm q)
\equiv \langle \alpha,\bm k|\rho^{\rm eff}_{\bm q+\bm G}|\beta,\bm k+\bm q\rangle~.
\end{equation}
To leading adiabatic order $\rho^{\rm eff}_{\bm Q}$ reduces to the usual projected density $P_+e^{i\bm Q\cdot\bm r}P_+$; finite-$J$ corrections from the SW dressing can be incorporated systematically as additional contributions to $F^{\alpha\beta}_{\bm G}$. For notational simplicity we will write $\rho_{\bm Q}$ for the dressed projected density $\rho^{\rm eff}_{\bm Q}$ throughout this section. Hermiticity $\rho_{-\bm q-\bm G}=\rho^\dagger_{\bm q+\bm G}$ follows from
\begin{equation}
F^{\beta\alpha}_{-\bm G}(\bm k+\bm q;-\bm q)=\Big(F^{\alpha\beta}_{\bm G}(\bm k;\bm q)\Big)^*~.
\end{equation}

When periodic modulations, including $\delta b$, scalar potentials, and non-adiabatic $1/J$ terms, are weak compared to $\omega_c$ or when a small Landau-level truncation is controlled, the form factors reduce to analytic Landau-level matrix elements dressed by miniband eigenvectors. Expanding $|\alpha,\bm k\rangle=\sum_{n,a}u^\alpha_{na}(\bm k)\,|n,a;\bm k\rangle$ (Landau level $n$ and internal magnetic-sublattice index $a=1,\dots,p$) gives
\begin{equation}
F^{\alpha\beta}_{\bm G}(\bm k;\bm q)=\sum_{n_1,n_2}\sum_{a_1,a_2} u^{\alpha *}_{n_1a_1}(\bm k)u^{\beta}_{n_2a_2}(\bm k+\bm q)
\langle n_1,a_1;\bm k|e^{i(\bm q+\bm G)\cdot\bm r}|n_2,a_2;\bm k+\bm q\rangle~,
\end{equation}
where the remaining matrix element is known analytically. Here we have written explicitly the leading adiabatic contribution in which $\rho^{\rm eff}_{\bm Q}\to P_+e^{i\bm Q\cdot\bm r}P_+$; the finite-$J$ SW dressing generates additional corrections $\delta F^{\alpha\beta}_{\bm G}$ that can be included systematically order-by-order in $1/J$. This provides a practical starting point for mostly analytic evaluation of response functions in the presence of moir\'e Umklapp mixing.

\subsection{Projected interactions and Umklapp-resolved matrix}

Projecting a two-body interaction into the retained minibands yields
\begin{equation}\label{eq:Hint_GG}
H_{\rm int}=\frac{1}{2A}\sum_{\bm q\in{\rm MBZ}}\sum_{\bm G,\bm G'\in\mathcal G}V^C_{\bm G\bm G'}(\bm q)\rho_{-\bm q-\bm G}\,\rho_{\bm q+\bm G'}~.
\end{equation}
For a translationally invariant two-body interaction $V_C(\bm r-\bm r')$,
\begin{equation}
V^C_{\bm G\bm G'}(\bm q)=V_C(\bm q+\bm G) \, \delta_{\bm G\bm G'}~.
\end{equation}

\paragraph*{Approximation scheme.} The form factors $F^{\alpha\beta}_{\bm G}$ are matrix elements of the dressed density operator within the chosen single-particle truncation and to the order retained in the SW dressing. To obtain interacting response we use a TDHF/RPA resummation in the Umklapp basis. This captures mode folding and oscillator-strength transfer due to moir\'e Umklapp mixing; quantitative dispersions may receive corrections beyond RPA.

The bare noninteracting polarization is a matrix in Umklapp indices,
\begin{align}\label{eq:PiGG}
\Pi_{\bm G\bm G'}(\bm q,\omega) &= \frac{1}{A}\sum_{\bm k\in{\rm MBZ}} \sum_{\alpha,\beta} \frac{n_F(\varepsilon_{\alpha,\bm k})-n_F(\varepsilon_{\beta,\bm k+\bm q})}{\omega+\varepsilon_{\alpha,\bm k}-\varepsilon_{\beta,\bm k+\bm q}+i0^+} \; F^{\alpha\beta}_{\bm G}(\bm k;\bm q)\; \Big(F^{\alpha\beta}_{\bm G'}(\bm k;\bm q)\Big)^{\!*}~.
\end{align}

The retarded density response matrix $\bm\chi(\bm q,\omega)$ obeys the Dyson equation
\begin{equation}
\bm\chi(\bm q,\omega) = \bm\Pi(\bm q,\omega) +\bm\Pi(\bm q,\omega)\,\bm V_C(\bm q)\,\bm\chi(\bm q,\omega)~,
\end{equation}
whose solution is the generalized matrix RPA form
\begin{equation}\label{eq:chiRPA}
\bm\chi_{\rm RPA}(\bm q,\omega) = \big[\bm 1-\bm\Pi(\bm q,\omega)\bm V_C(\bm q)\big]^{-1}\bm\Pi(\bm q,\omega) = \big[\bm\Pi^{-1}(\bm q,\omega)-\bm V_C(\bm q)\big]^{-1}~.
\end{equation}
Collective modes correspond to poles of $\bm\chi_{\rm RPA}$, equivalently
\begin{equation}\label{eq:collectiveDet}
\det \big[\bm 1-\bm\Pi(\bm q,\omega)\bm V_C(\bm q)\big]=0~.
\end{equation}

\begin{figure}[t]
  \centering
  \includegraphics[width=0.6\linewidth]{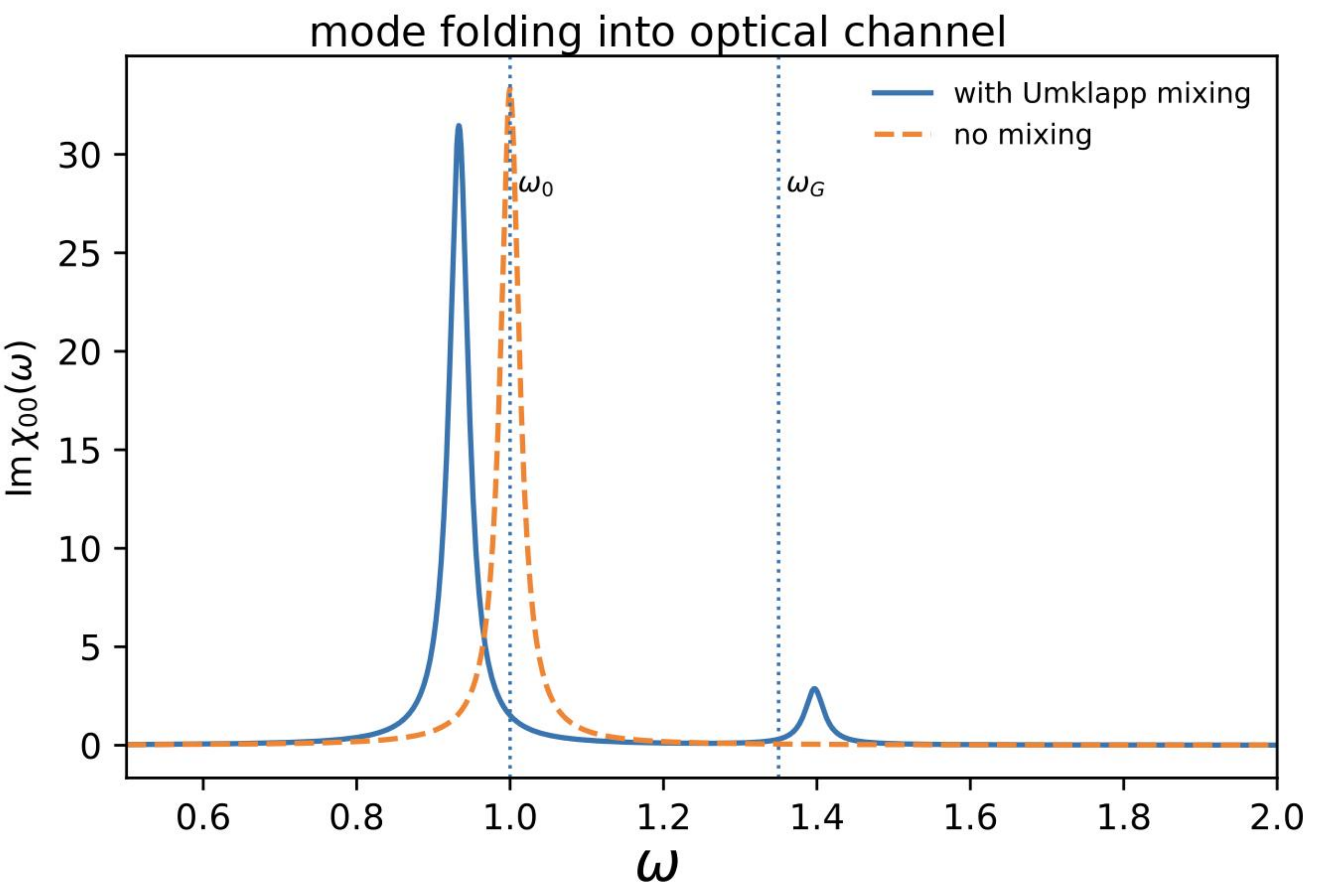}
  \caption{Two-sector Umklapp mixing example: coupling between $\bm G=0$ and $\bm G\neq 0$ density sectors transfers spectral weight, making a finite-momentum collective mode optically active (``mode folding'') in the $\bm G=0$ channel.}
  \label{Fig3}
\end{figure}

\paragraph*{Optical channel and Umklapp leakage.} Separating the uniform $\bm G=0$ component from $\bm G\neq 0$ components, one can integrate out the Umklapp sector to obtain an effective polarization for the optical response.
Let $r$ denote the block of Umklapp indices with $\bm G\neq 0$, and assume $\bm V_C$ is diagonal in $\bm G$.
Define
\begin{equation}\label{eq:Pi00_eff}
\Pi_{00}^{\rm eff}(\bm q,\omega) \equiv \Pi_{00}
+ \Pi_{0r} \bm V^C_r \big[\bm 1-\bm\Pi_{rr}\bm V^C_r\big]^{-1} \Pi_{r0}~,
\end{equation}
which resums repeated Umklapp scattering.
Then the optical response takes the RPA form
\begin{equation}\label{eq:chi00_eff}
\chi_{00}(\bm q,\omega)
= \frac{\Pi_{00}^{\rm eff}(\bm q,\omega)}{1-V_C(\bm q)\,\Pi_{00}^{\rm eff}(\bm q,\omega)}~,
\end{equation}
which makes the key consequence explicit: collective modes that primarily reside at finite $\bm G$ can acquire spectral weight in the uniform channel through the Umklapp mixing encoded in $\Pi_{0r}$, producing ``shadow'' poles in $\chi_{00}$, see Fig.~\ref{Fig3}.

Importantly, the appearance of folded (Umklapp) poles in the $\bm G=0$ channel is a generic consequence of moir\'e periodicity and the matrix structure of the density operator; it does not rely on any special microscopic fine tuning.

\section{Effective Field Theory of a Skyrmion Crystal from the Microscopic Model}
\label{sec:effective_field_theory}

In this section we derive a long-wavelength effective field theory for a skyrmion crystal starting from a microscopic continuum description of electrons coupled to a slowly varying pseudospin texture. We focus on the universal collective degrees of freedom of a periodic skyrmion texture, its magneto-elastic deformations (phonons) and, more generally, its topological defects, rather than constructing a full nonlinear sigma model for $\hat{\bm n}$. The EFT is organized to make explicit which steps are controlled and which inputs are many-body assumptions. A central theme is the separation between parity-odd and parity-even data: the parity-odd kinematics are fixed by quantized Hall response, while parity-even elastic coefficients are nonuniversal and can be obtained by microscopic matching.

\paragraph*{Microscopic-to-EFT logic.} The construction proceeds in two conceptual steps. 
First, a controlled operator-level Schrieffer-Wolff reduction in $1/J$ integrates out the high-energy pseudospin branch and yields an effective single-branch theory together with systematically dressed physical operators, e.g. density, current, etc. This defines the interacting projected low-energy theory once interactions are expressed in terms of the dressed density operators. Second, to obtain a local long-wavelength EFT for a skyrmion crystal, we further assume that the projected interacting system realizes a fully gapped Hall phase. Under this assumption, the gapped electronic sector can be integrated out, and gauge invariance implies a Chern-Simons term with a quantized coefficient. Expanding about a periodic skyrmion-crystal saddle point then yields a magneto-elastic theory for crystal deformations with a universal parity-odd Berry-phase term fixed by the Hall response.

\paragraph*{Scope and inputs.} Throughout this section $J(\bm r)\hat{\bm n}(\bm r)$ denotes a generic slowly varying pseudospin splitting field that separates two local branches by an energy $\sim 2J(\bm r)$. Depending on the microscopic setting, this splitting can arise from single-particle moir\'e terms and/or from interaction-driven self-consistent order. The controlled SW step requires only that $J$ is the largest local scale and that the texture varies slowly; it does not by itself determine which correlated phase is realized at a given filling. The gapped-Hall assumption enters only when we write the topological Chern-Simons response and the resulting universal magneto-elastic EFT.

\paragraph*{Gauge-field notation.} We distinguish three gauge fields. The external electromagnetic probe is $A_\mu$ with $e$ kept explicit. A local $SU(2)$ rotation $U(\bm r,\tau)$ into the texture-aligned frame generates an emergent non-Abelian gauge field $\mathcal A_\mu \equiv iU^\dagger \, \partial_\mu U$. Upon projection to a single branch, the diagonal component defines an emergent Abelian Berry connection $a_\mu[\hat{\bm n}]$. The projected electronic theory couples to the combination
\begin{equation}\label{eq:Bmu_def}
B_\mu \equiv eA_\mu + a_\mu[\hat{\bm n}]~,
\end{equation}
and the parity-odd response of a gapped Hall phase is a functional of $B_\mu$.

\paragraph*{Relation to the skyrmion-Chern-band Hamiltonian.} In Sec.~\ref{sec:skyrmion_chern_band} we introduced a two-component moir\'e Hamiltonian with a static splitting field $J(\bm r)\,\hat{\bm n}(\bm r)\cdot\bm\sigma$ that encodes a periodic pseudospin texture and generates a skyrmion Chern band in the adiabatic limit. Here we promote the same field to a slow collective degree of freedom and derive its long-wavelength theory. The emergent gauge structure and the controlled $1/J$ projection rely only on the existence of a large local branch splitting and on slow space-time variation of the local eigenbasis; they do not depend on the microscopic origin of $J\hat{\bm n}$. Consequently, moir\'e single-particle terms and Coulomb interactions determine the static saddle point $(J_0(\bm r),\hat{\bm n}_0(\bm r))$ and renormalize parity-even coefficients, while the parity-odd response is fixed by the quantized Hall coefficient $\kappa$ as long as the many-body gap remains open.

\subsection{Microscopic action and collective texture field}

We work in Euclidean spacetime $x\equiv(\tau,\bm r)$ with $d^3x=d\tau\,d^2r$ and $\tau\in[0,\beta]$. (In real time we write $d^3x=dt\,d^2r$.) In the presence of an external probe $A_\mu$, the partition function is
\begin{equation}
Z[A] = \int \mathcal D\psi^\dagger\,\mathcal D\psi\; e^{-S_E[\psi,A]}~.
\end{equation}
We start from a two-component fermion $\psi$ (layer/pseudospin) with Coulomb interactions and allow for a slowly varying pseudospin splitting field. A convenient Euclidean action in the presence of an external probe $A_\mu$ is
\begin{equation}
S_E[\psi,A]
= \int d^3x\;\psi^\dagger\!\left(D_\tau+\frac{\Pi_i^2}{2m}+V(\bm r)\right)\!\psi
+\frac{g_{\text{in}}}{2}\int d^3x\;\big(\bm O(x)\big)^2~,
\label{eq:S_micro_E_recap}
\end{equation}
where $D_\tau \equiv \partial_\tau - i e A_\tau$, $\Pi_i \equiv -i\partial_i - e A_i$, $\bm O(x)\equiv \psi^\dagger(x)\,\bm\sigma\,\psi(x)$ is the pseudospin density, and $g_{\text{in}}$ denotes an effective short-distance interaction in the pseudospin channel. We emphasize that the Hubbard-Stratonovich (HS) construction below is used as a convenient device to promote the splitting field to a collective variable and to organize its coupling to gauge fields; the universal parity-odd terms derived later depend only on the existence of a gapped Hall phase and $U(1)$ charge conservation, not on the detailed microscopic interaction model.

Introducing an HS field $\bm M$,
\begin{equation}\label{eq:HS_repulsive_i}
\exp\!\left[-\frac{g_{\text{in}}}{2}\int d^3x\;\bm O^2\right] \propto \int \mathcal D\bm M\; \exp\!\left[-\int d^3x\left(\frac{\bm M^2}{2g_{\text{in}}}-i \bm M\cdot \bm O\right)\right]~,
\end{equation}
we parameterize the resulting splitting field as
\begin{equation}\label{eq:def_exchange_field}
\bm M(x)\equiv i J(x)\hat{\bm n}(x)~,\qquad J(x)\ge 0~,\qquad \hat{\bm n}^2=1~.
\end{equation}
We may view $\bm M=iJ\hat{\bm n}$ as a convenient contour rotation (analytic continuation) that makes the fermion coupling $-J\hat{\bm n}\cdot\bm\sigma$ real; the universal parity-odd response derived below does not depend on the HS contour choice. The HS-decoupled action reads
\begin{align}\label{eq:S_HS}
\begin{split}
S_E[\psi,J,\hat{\bm n};A]=\int d^3x\;\psi^\dagger &\left(D_\tau +\frac{\Pi_i^2}{2m}+V(\bm r)- J(x) \hat{\bm n}(x)\cdot\bm\sigma\right)\psi \\
&-\int d^3x\;\frac{J(x)^2}{2g_{\text{in}}} +S_{{\rm bare},E}[\hat{\bm n}]~.    
\end{split}
\end{align}
Integrating out $\psi$ gives the order-parameter effective action
\begin{equation}\label{eq:Seff_trlog}
S_{\rm eff}[J,\hat{\bm n},A]
= -\!\int d^3x\;\frac{J(x)^2}{2g_{\text{in}}}+S_{{\rm bare},E}[\hat{\bm n}]
-\Tr\ln\!\left[D_\tau+\frac{\Pi_i^2}{2m}+V(\bm r)-J(x)\,\hat{\bm n}(x)\cdot\bm\sigma\right]~.
\end{equation}
which provides a formal starting point for deriving parity-even gradient terms and mean-field/RPA fluctuation kernels for the texture. In the remainder of this section, however, we keep the fermionic fields and instead use the controlled Schrieffer-Wolff reduction to define the projected interacting theory and its dressed operators; only after assuming that this projected many-body problem forms a gapped Hall phase do we integrate out the gapped electronic sector to obtain the universal Chern-Simons term. A skyrmion-crystal configuration corresponds to a static periodic saddle point
\begin{equation}
J(x)\to J_0(\bm r)~, \qquad \hat{\bm n}(x)\to \hat{\bm n}_0(\bm r)    
\end{equation}
which we assume exists in the parameter regime of interest.

\paragraph*{Gaussian fluctuations.} Expanding about a saddle point $J\hat{\bm n}=J_0\hat{\bm n}_0+\delta \bm M$, yields the standard quadratic fluctuation kernel
\begin{equation}
S_{\rm eff}=S_{\rm eff}^{(0)}+\frac12\int d^3x\,d^3x'\;\delta M_a(x)\,K_{ab}(x,x')\,\delta M_b(x')+\cdots~,
\end{equation}
with the ``bare$+$bubble'' structure
\begin{equation}\label{eq:kernel_def}
K_{ab}(x,x')=-\frac{1}{g_{\text{in}}}\delta_{ab}\delta^{(3)}(x-x')+\Pi_{ab}(x,x')+K^{\rm(bare)}_{ab}(x,x')~,
\end{equation}
\begin{equation}\label{eq:Pi_bubble}
\Pi_{ab}(x,x')=\Tr\left[G_0(x,x')\sigma_a\,G_0(x',x)\sigma_b\right]~,\qquad
G_0\equiv G_{J_0\hat{\bm n}_0}~,
\end{equation}
where $\Tr$ is the trace over pseudospin indices and $G_0$ is the fermion Green's function. For a periodic saddle point, the diagonalization of $K$ is naturally organized in a magnetic-Bloch basis, and collective mode dispersions follow from $\det K(\omega,\bm q)=0$ after analytic continuation.

\subsection{Local \texorpdfstring{$SU(2)$}{SU(2)} rotation and emergent gauge structure}

To expose the geometric structure, we rotate to a local frame aligned with the texture. Let $U(\bm r,\tau)\in SU(2)$ satisfy
\begin{equation}\label{eq:su2_rotation}
U^\dagger(\bm r,\tau)\,\hat{\bm n}(\bm r,\tau)\cdot\bm\sigma\,U(\bm r,\tau)=\sigma_z~, 
\qquad \psi(\bm r,\tau)=U(\bm r,\tau)\,\chi(\bm r,\tau)~.
\end{equation}
The spacetime variation of $U$ generates an emergent non-Abelian gauge field
\begin{equation}\label{eq:nonabelian_gauge}
\mathcal A_\mu(\bm r,\tau)\equiv i U^\dagger \partial_\mu U~, \qquad \mu\in\{\tau,x,y\}~,
\end{equation}
so that derivatives acting on $\chi$ acquire $\mathcal A_\mu$. In the rotated frame the action becomes
\begin{equation}\label{eq:rotated_action}
S_E[\chi,\mathcal A;A]=\int d^3x\;\chi^\dagger\!\left[D_\tau-i\mathcal A_\tau+\frac{\big(\bm p-e\bm A-\bm{\mathcal A}\big)^2}{2m}+V(\bm r)-J\sigma_z\right]\!\chi +S_{{\rm int},E}[\chi]+\cdots~,
\end{equation}
where $\bm{\mathcal A}\equiv(\mathcal A_x,\mathcal A_y)$, and the ellipsis denotes the $-J^2/(2g_{\text{in}})$ and $S_{\rm bare}$ terms already included in~\eqref{eq:S_HS}. Eq.~\eqref{eq:rotated_action} is exact: the emergent gauge structure is generated by a change of local basis and does not assume adiabaticity.

The residual freedom 
\begin{equation}
U\to U e^{i\varphi\sigma_z/2}~,   
\end{equation}
leaves the alignment condition invariant and implies a $U(1)$ gauge redundancy in the adiabatic subspace. Denote by $a_\mu$ the diagonal Abelian component of $\mathcal A_\mu$ in the low-energy branch; equivalently, in a $\mathbb{CP}^1$ parameterization $\hat{\bm n}=z^\dagger\bm\sigma z$ with a $\mathbb{CP}^1$ spinor $z(\bm r,\tau)$ and $z^\dagger z=1$, one may write
\begin{equation}\label{eq:berry_connection}
a_\mu(\bm r,\tau)\equiv -i z^\dagger\partial_\mu z~,
\end{equation}
which shifts under $z\to e^{i\phi}z$ as $a_\mu\to a_\mu+\partial_\mu\phi$.

\subsection{Controlled projection to a single branch and the projected interacting theory}

We now implement the controlled step: integrate out the high-energy pseudospin branch in the regime of a large local splitting $2J(\bm r)$. Formally, we perform an operator-level SW reduction in powers of $1/J$, which is controlled when $J$ exceeds the moir\'e kinetic scale and the texture varies slowly on the moir\'e scale. The SW procedure applies equally in the presence of interactions: it block-diagonalizes the microscopic Hamiltonian and, crucially, induces a corresponding dressing of physical operators. In the interacting many-body problem this expansion is controlled provided $J$ exceeds the typical energy scales that couple the two branches (moir\'e kinetic scale and interaction-induced inter-branch matrix elements at the moir\'e length scale).

At the level of the fermion action~\eqref{eq:rotated_action}, let $P$ and $Q$ project onto the low- and high-energy branches in the rotated frame. The SW expansion yields an effective low-energy operator in the $P$ subspace,
\begin{equation}\label{eq:SW_expansion}
\mathcal H_{\rm eff}=P\mathcal V P - P\mathcal V Q\,\frac{1}{2J}\,Q\mathcal V P +\mathcal O(J^{-2})~,
\end{equation}
where $\mathcal V$ contains the kinetic energy, $V(\bm r)$, and the gauge couplings generated by $\mathcal A_\mu$.

To leading order, the low-energy dynamics reduces to a single fermion field $\chi$ minimally coupled to the projected Abelian Berry connection $a_\mu[\hat{\bm n}]$ (the diagonal component of $\mathcal A_\mu$ in the low-energy branch),
\begin{equation}\label{eq:adiabatic_action}
S_{\rm 1b}[\chi;A,\hat{\bm n}] =\int d^3x\;\chi^\dagger\!\left[\partial_\tau - i(eA_\tau+a_\tau)+\frac{\big(\bm p-e\bm A-\bm a\big)^2}{2m}+V_{\rm eff}(\bm r)\right]\!\chi + S_{{\rm int},E}[\chi]~,
\end{equation}
where $V_{\rm eff}$ collects smooth scalar terms including geometric scalar terms generated by texture gradients, and the $\mathcal O(J^{-1})$ corrections are fixed by the real-space quantum geometric tensor of the local adiabatic state, as derived in Sec.~\ref{sec:controlled_expansions}.

\paragraph*{Operator dressing and the interacting projected theory.} The same SW unitary that produces~\eqref{eq:adiabatic_action} must be applied to physical operators. In particular, the low-energy charge density is the dressed operator $\rho^{\rm eff}\equiv P\,e^{S}\rho\,e^{-S}P$. For density-density interactions, this defines the projected interacting Hamiltonian in the standard form but with dressed density operators and corresponding form factors. This point is essential for the many-body interpretation: the controlled SW reduction provides a microscopic definition of the interacting projected theory; it does not by itself determine which correlated phase that theory realizes at a given filling. For example, a density-density interaction takes the standard form
$S_{{\rm int},E}=\frac12\int_{\tau,\bm q}V_C(\bm q)\,\rho^{\rm eff}_{-\bm q}(\tau)\rho^{\rm eff}_{\bm q}(\tau)$, with $\rho^{\rm eff}$ the SW-dressed projected density.

\subsection{Skyrmion current and emergent flux}

Define the field strength of the emergent Abelian connection,
\begin{equation}
f_{\mu\nu}\equiv \partial_\mu a_\nu-\partial_\nu a_\mu~.
\end{equation}
The topological skyrmion current of the texture is
\begin{equation}\label{eq:sk_current_def}
J_\mu^{\rm sk} \equiv \frac{1}{8\pi}\varepsilon_{\mu\nu\lambda}\,
\hat{\bm n}\cdot(\partial_\nu\hat{\bm n}\times\partial_\lambda\hat{\bm n})~,
\end{equation}
and the $\mathbb{CP}^1$ identity relates it directly to $a_\mu$,
\begin{equation}\label{eq:sk_current_flux}
J_\mu^{\rm sk} =\frac{1}{4\pi}\varepsilon_{\mu\nu\lambda}f_{\nu\lambda}
=\frac{1}{2\pi}\varepsilon_{\mu\nu\lambda}\partial_\nu a_\lambda~.
\end{equation}
In particular, the static skyrmion density equals the emergent magnetic flux density,
\begin{equation}\label{eq:sk_density_bfield}
\rho_{\rm sk}(\bm r)\equiv J_\tau^{\rm sk} =\frac{1}{4\pi}\hat{\bm n}\cdot(\partial_x\hat{\bm n}\times\partial_y\hat{\bm n}) =\frac{1}{2\pi}b(\bm r)~,\qquad b(\bm r)\equiv f_{xy}(\bm r)~.
\end{equation}
For a periodic texture, the total emergent flux per unit cell is quantized,
\begin{equation}\label{eq:flux_quantization}
\int_{\rm u.c.} d^2r\;b(\bm r)=2\pi Q~,\qquad Q\in\mathbb Z~.
\end{equation}

\subsection{Chern-Simons term of a gapped Hall phase}

We now invoke the many-body assumption needed for a local EFT. We assume that the projected interacting theory defined after the controlled SW reduction forms a fully gapped Hall phase. At energies well below the many-body gap, locality and gauge invariance imply that the leading parity-odd response to the combined gauge field $B_\mu$ defined in Eq.~\eqref{eq:Bmu_def} is a Chern-Simons term. If the electronic sector is gapless, e.g. composite Fermi liquid, the Chern-Simons reduction must be replaced by a nonlocal dynamical polarization kernel. Smooth $1/J$ corrections can renormalize parity-even coefficients but cannot change the quantized Chern-Simons coefficient as long as the bulk gap and $U(1)$ charge conservation remain intact. 

It is convenient to define the dimensionless Hall coefficient
\begin{equation}
\kappa \equiv \frac{2\pi}{e^2}\sigma_{xy}\in\mathbb Q~,
\end{equation}
so that the universal topological response takes the form
\begin{equation}\label{eq:CS_general}
S_{\rm top}[B] =\frac{\kappa}{4\pi}\int d^3x\;\varepsilon^{\mu\nu\rho}B_\mu\,\partial_\nu B_\rho~.
\end{equation}
For a filled integer Chern band, $\kappa=C$ equals the total Chern number of the occupied bands. For an Abelian fractional Chern insulator described by a $K$-matrix theory~\cite{1992IJMPB...6.1711W,WenZeePRB1992KMatrix}, $\kappa=t^T K^{-1}t$ is the many-body Hall response. Expanding in $A_\mu$ and $a_\mu$ yields the mixed coupling
\begin{equation}\label{eq:A_Jsk}
S_{A\text{-}{\rm sk}}=\frac{\kappa e}{2\pi}\int A\,da =\kappa e\int d^3x\;A_\mu J_\mu^{\rm sk}~,
\end{equation}
where the second equality follows from~\eqref{eq:sk_current_flux}. Therefore, a texture event carrying skyrmion number $Q$ binds electric charge
\begin{equation}\label{eq:sk_charge}
Q_{\rm el}[Q]=\kappa e Q~.
\end{equation}

From Eq.~\eqref{eq:CS_general} we obtain three universal implications for the coupled electromagnetic-texture response:

\paragraph*{1 Electromagnetic response and texture pumping.} The topological response to the combined gauge field $B_\mu \equiv eA_\mu+a_\mu[\hat{\bm n}]$ implies an electromagnetic current. Varying with respect to the electromagnetic probe gives the physical charge current
\begin{equation}\label{eq:jmu_from_CS}
j^\mu_{\rm el}\equiv \frac{\delta S_{\rm top}}{\delta A_\mu}
=\frac{\kappa e}{2\pi}\varepsilon^{\mu\nu\lambda}\partial_\nu B_\lambda
=\frac{\kappa e^2}{2\pi}\varepsilon^{\mu\nu\lambda}\partial_\nu A_\lambda
+\frac{\kappa e}{2\pi}\varepsilon^{\mu\nu\lambda}\partial_\nu a_\lambda~.
\end{equation}
Using the identity for the skyrmion topological current,
\begin{equation}
J^\mu_{\rm sk}\equiv \frac{1}{2\pi}\varepsilon^{\mu\nu\lambda}\partial_\nu a_\lambda~,
\end{equation}
Eq.~\eqref{eq:jmu_from_CS} can be written in the transparent form
\begin{equation}\label{eq:jmu_pumping_split}
j^\mu_{\rm el}=\frac{\kappa e^2}{2\pi}\varepsilon^{\mu\nu\lambda}\partial_\nu A_\lambda +\kappa e J^\mu_{\rm sk}~.
\end{equation}

From this point on we analytically continue to real time ($\tau\to it$) and write $A_0$ for the temporal component; the Chern-Simons response is understood as the real-time effective action governing low-frequency dynamics and response. In components, defining
$B_{\rm ext}\equiv \varepsilon^{ij}\partial_i A_j$,
$b\equiv \varepsilon^{ij}\partial_i a_j$,
$E_i\equiv -\partial_t A_i+\partial_i A_0$, and
$\mathcal E^{a}_i\equiv -\partial_t a_i+\partial_i a_0$,
one obtains
\begin{align}
\rho_{\rm el}\equiv j^0_{\rm el}
&= \frac{\kappa e}{2\pi}\big(eB_{\rm ext}+b\big)
= \frac{\kappa e^2}{2\pi}B_{\rm ext}+\kappa e \rho_{\rm sk}~,
\label{eq:rho_response_polished}\\
j^i_{\rm el}
&= \frac{\kappa e}{2\pi}\varepsilon^{ij}\big(eE_j+\mathcal E^a_j\big) = \frac{\kappa e^2}{2\pi}\varepsilon^{ij}E_j+\kappa e J^i_{\rm sk}~.
\label{eq:ji_response_polished}
\end{align}
which show that a time-dependent texture
($\mathcal E_j^a\neq 0$, equivalently $J^i_{\rm sk}\neq 0$) pumps electrical current even at $B_{\rm ext}=0$, providing a direct route from skyrmion dynamics to optical and transport signatures.

\paragraph*{2. A Hopf term for the texture and skyrmion statistics.} Setting $A_\mu=0$ in~\eqref{eq:CS_general} leaves a Hopf term for $\hat{\bm n}$ written in terms of $a_\mu[\hat{\bm n}]$:
\begin{equation}\label{eq:Hopf}
S_{\rm Hopf}[\hat{\bm n}]=\frac{\kappa}{4\pi}\int d^3x\;\varepsilon^{\mu\nu\rho}a_\mu[\hat{\bm n}]\,\partial_\nu a_\rho[\hat{\bm n}]~,
\end{equation}
which fixes the Berry phase structure of skyrmion dynamics and, in fractional quantum Hall phases, the anyonic statistics of skyrmions.

\paragraph*{3. The nonuniversal stiffness functional.} Parity-even terms generated by integrating out the gapped electrons produce a nonuniversal energy functional $E[\hat{\bm n}]$ and higher-gradient corrections,
\begin{equation}
S_{\rm even}[\hat{\bm n}]=\int d t\;E[\hat{\bm n}]+\cdots~,\qquad
E[\hat{\bm n}]=\int d^2r\left[\frac{\rho_s}{2}(\partial_i\hat{\bm n})^2+V_{\rm ani}(\hat{\bm n})+\cdots\right]~,
\end{equation}
whose minimum may be a periodic skyrmion crystal configuration. Microscopically, these coefficients can be computed in controlled limits (e.g. large-$J$ expansions in mean-field settings) or obtained by matching to microscopic energetics of the projected theory.

\subsection{Skyrmion crystal: magneto-elastic effective field theory for phonons}

We now specialize to a skyrmion-crystal ground state and derive the universal long-wavelength theory of its translational deformations. Assume that the parity-even functional admits a static periodic minimum $\hat{\bm n}_0(\bm r)$ with unit-cell area $A_M$ and integer skyrmion number per cell $Q$,
\begin{equation}
\hat{\bm n}_0(\bm r+\bm R)=\hat{\bm n}_0(\bm r)~,\qquad
\int_{A_M} d^2r\;\rho_{\rm sk,0}(\bm r)=Q~,\qquad 
\bar\rho_{\rm sk}\equiv Q/A_M~.
\end{equation}
The universal low-energy degrees of freedom associated with broken translations are described by a displacement field $\bm u(\bm r,t)$, implemented as a slow deformation of the background texture,
\begin{equation}
\hat{\bm n}(\bm r,t)=\hat{\bm n}_0\big(\bm r-\bm u(\bm r,t)\big)+\cdots~,
\end{equation}
possibly supplemented by additional internal soft modes, e.g. global phase/rotation, if present. In this work we focus on the universal translational phonons.

\paragraph*{Parity-even nonuniversal elasticity.} Expanding the parity-even functional about $\hat{\bm n}_0$ yields the elastic energy
\begin{equation}\label{eq:elastic_action}
S_{\rm el}[\bm u] =\frac12\int dt\,d^2r\;C_{ijkl}\,u_{ij}u_{kl}+\cdots~,
\qquad u_{ij}\equiv \frac12(\partial_i u_j+\partial_j u_i)~,
\end{equation}
with elastic moduli $C_{ijkl}$. These coefficients are nonuniversal and may be obtained by matching to a microscopic energy functional for the projected theory and receive controlled $1/J$ renormalizations.

\paragraph*{Microscopic matching of elastic moduli.} The parity-even part of the fermion determinant fixes the static energy functional $E[\hat{\bm n}]$ and thereby the elastic moduli of a skyrmion-crystal saddle point $\hat{\bm n}_0(\bm r)$. A convenient definition is uniform-strain matching:
\begin{equation}\label{eq:elastic_matching}
E\!\left[\hat{\bm n}_0 \big((\mathbf 1-\bm\varepsilon)\bm r\big)\right]
=E_0+\frac{A}{2} C_{ijkl}\,\varepsilon_{ij}\varepsilon_{kl}+\cdots~,
\end{equation}
where $\bm\varepsilon$ is a constant strain tensor and $A$ is the sample area.
In practice, $E[\hat{\bm n}]$ may be evaluated from a controlled large-$J$ derivative expansion including the leading $1/J$ corrections or from a microscopic energy functional, e.g., Hartree-Fock, for the underlying skyrmion Chern band model, while the parity-odd response terms are fixed by the quantized coefficient $\kappa$ as long as the many-body gap remains open.

\paragraph*{Parity-odd universal Berry-phase term.} By contrast, the parity-odd term is fixed by the quantized Hall response of the gapped electronic background. Setting $A_\mu=0$ for the moment, the Chern-Simons term for $B_\mu=eA_\mu+a_\mu$ implies a universal topological action for $a_\mu$,
\begin{equation}\label{eq:ada_term}
S_{a}=\frac{\kappa}{4\pi}\int d^3x\;\varepsilon^{\mu\nu\lambda}a_\mu\,\partial_\nu a_\lambda~.
\end{equation}
For a skyrmion crystal, the low-energy distortions are captured by the displacement field $\bm u(\bm r,t)$ via the translated texture $\hat{\bm n}(\bm r,t)=\hat{\bm n}_0\big(\bm r-\bm u(\bm r,t)\big)+\cdots$. Accordingly, the emergent connection is pulled back as
\begin{equation}
a_\mu(\bm r,t)=a_{\mu,0}\big(\bm r-\bm u(\bm r,t)\big)+\cdots~,
\end{equation}
where $a_{\mu,0}\equiv a_\mu[\hat{\bm n}_0]$. Substituting into~\eqref{eq:ada_term} and expanding to leading order in $\partial_t\bm u$ yields the Berry-phase term
\begin{equation}\label{eq:berry_u}
S_{\rm B}[\bm u] =\frac{\Gamma}{2}\int dt\,d^2r\;\varepsilon_{ij} u_i\,\partial_t u_j+\cdots~,  \qquad \Gamma\equiv \kappa \bar\rho_{\rm sk}~,
\end{equation}
which is universal as long as the many-body gap remains open and $\bar\rho_{\rm sk}\equiv Q/A_M$ is the mean skyrmion density of the crystal with $Q$ the skyrmion number per moir\'e unit cell of area $A_M$. It implies that $u_x$ and $u_y$ form a canonically conjugate pair. The canonical momentum density is
\begin{equation}
\pi_i(\bm r)\equiv \frac{\delta \mathcal L}{\delta(\partial_t u_i)} =\frac{\Gamma}{2}\varepsilon_{ji}u_j(\bm r)~,
\end{equation}
and canonical quantization gives the local noncommutative algebra
\begin{equation}\label{eq:noncomm_u_polished}
[u_x(\bm r),u_y(\bm r')]=\frac{i}{\Gamma}\,\delta(\bm r-\bm r')+\cdots~,
\end{equation}
which is the direct skyrmion-crystal analogue of guiding-center noncommutativity in a Landau level. Consequently, in the clean unpinned limit the two broken translations pair into a single type-B Goldstone mode.

\paragraph*{Coupling to electromagnetism and density modulation.} From~\eqref{eq:A_Jsk}, crystal deformations couple to $A_\mu$ via the induced change in skyrmion current:
\begin{equation}\label{eq:EM_coupling_u}
S_{\rm em}[\bm u;A] = \kappa e\int d^3x\; A_\mu J^{\rm sk}_\mu[\hat{\bm n}_0(\bm r-\bm u)]~.
\end{equation}
In particular, smooth compressional deformations modulate the skyrmion density,
\begin{equation}
\delta \rho_{\rm sk}(\bm r,t)= -\bar\rho_{\rm sk} \nabla\cdot\bm u(\bm r,t)+\cdots~,
\end{equation}
and therefore the electric charge density,
\begin{equation}\label{eq:charge_modulation}
\delta \rho_{\rm el}(\bm r,t)=\kappa e\,\delta\rho_{\rm sk}(\bm r,t)
= -\kappa e \bar\rho_{\rm sk}\nabla\cdot\bm u(\bm r,t)+\cdots~,
\end{equation}
where the last step holds for smooth distortions.

\paragraph*{Long-wavelength effective field theory.} Collecting the universal contributions, the long-wavelength dynamics of a clean skyrmion crystal can be organized into a magneto-elastic effective action for the displacement field $\bm u(\bm r,t)$,
\begin{equation}\label{eq:SkX_EFT_final}
S_{\rm SkX}[\bm u;A] =\int d^3x\;\left[\frac{\Gamma}{2}\varepsilon_{ij}u_i\,\partial_t u_j-\frac{1}{2}C_{ijkl}\,u_{ij}\,u_{kl}\right] +\kappa e\int d^3x\;A_\mu J^\mu_{\rm sk}\left[\hat{\bm n}_0(\bm r-\bm u)\right]+\cdots~,
\end{equation}
where $u_{ij} \equiv \frac12(\partial_i u_j+\partial_j u_i)$ is the strain tensor, $C_{ijkl}$ are the elastic moduli, and
\begin{equation}
\Gamma\equiv \kappa \bar\rho_{\rm sk}~,\qquad \bar\rho_{\rm sk}\equiv Q/A_M~.
\end{equation}
The first-order Berry phase term is fixed by the quantized topological response coefficient $\kappa$ and the mean skyrmion density $\bar\rho_{\rm sk}$, and it renders $u_x$ and $u_y$ canonically conjugate. The ellipsis in Eq.~\eqref{eq:SkX_EFT_final} denotes higher-gradient elastic terms, explicit moir\'e pinning, and controlled non-adiabatic $1/J$ corrections. A conventional inertial term $\frac{\rho_u}{2} \dot u_i^2$ is symmetry allowed; it generates a high-frequency gapped partner mode and is subleading for $\omega \ll \Gamma/\rho_u$ so we neglect it in the long-wavelength theory. As long as the many-body gap of the underlying Hall state remains open, $\kappa$ is not renormalized by smooth $1/J$ corrections; such corrections instead renormalize parity-even coefficients, e.g.\ elastic moduli and higher-derivative terms. 

\paragraph*{Controlled finite-$J$ corrections.} Beyond strict adiabaticity, the Schrieffer-Wolff reduction generates a gauge-covariant correction $\delta\mathcal H_{1/J}$ to the single-branch Hamiltonian. If the interacting single-branch problem realizes a gapped Hall phase, integrating out the electrons yields an effective action of the form $S_{\rm eff}[A,\hat{\bm n}]=S_{\rm top}[B]+S_{\rm even}[B,\hat{\bm n}]+\cdots$ with $B_\mu=eA_\mu+a_\mu[\hat{\bm n}]$. The topological Chern-Simons coefficient $\kappa$ is quantized and cannot be renormalized by smooth $1/J$ corrections as long as the bulk gap and $U(1)$ charge conservation are preserved; by contrast, $\delta\mathcal H_{1/J}$ renormalizes only the parity-even functional $S_{\rm even}$, i.e., the elastic moduli and higher-derivative terms. A systematic cumulant expansion and matching procedure is discussed in Appendix~\ref{sec:map_1overJ_to_EFT}.

\begin{figure}[t]
  \centering
  \includegraphics[width=0.6\linewidth]{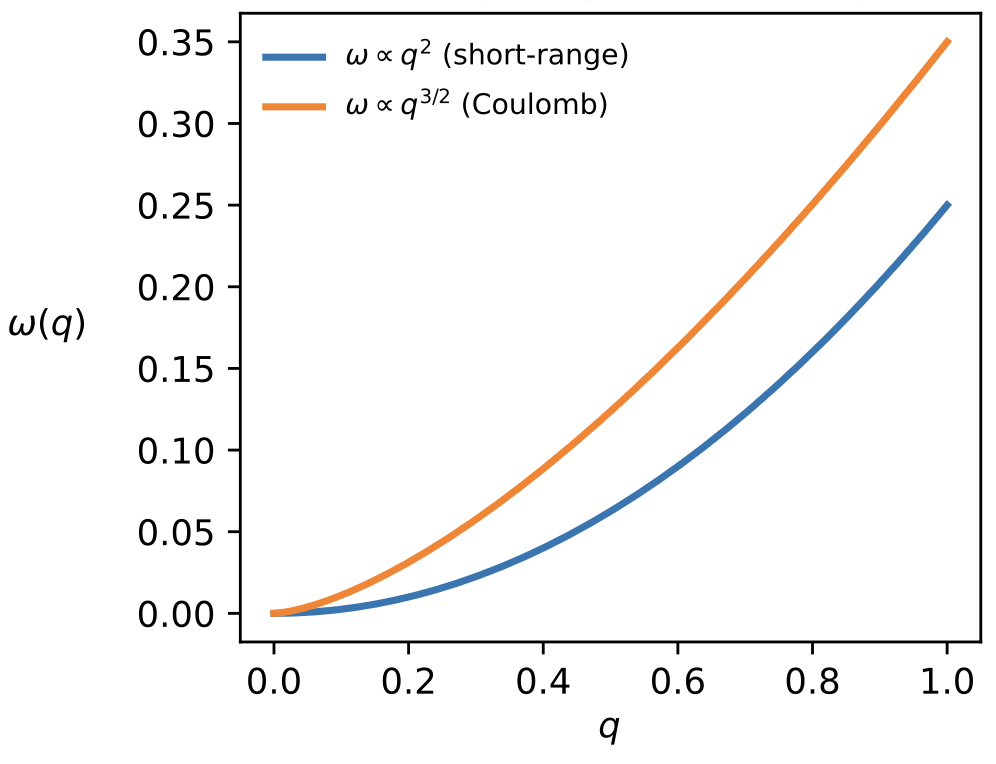}
  \caption{Skyrmion-crystal magnetophonon dispersion from the magneto-elastic effective field theory: $\omega\sim q^2$ for short-range elasticity, crossover to $\omega\sim q^{3/2}$ with long-range Coulomb interactions, and a pinning-induced gap under weak moir\'e pinning.}
  \label{dispersion}
\end{figure}

\paragraph*{Magnetophonon dispersion relation.} For an isotropic crystal with short-range elasticity,
\begin{equation}
C_{ijkl}=\lambda\,\delta_{ij}\delta_{kl} +\mu(\delta_{ik}\delta_{jl}+\delta_{il}\delta_{jk})~,
\end{equation}
and in the absence of pinning, the linearized equations of motion following from
Eq.~\eqref{eq:SkX_EFT_final} and inserting a plane-wave ansatz $\bm u= \bar{\bm u}\, e^{i(\bm q\cdot\bm r-\omega t)}$ yields a single type-B Goldstone mode with quadratic dispersion,
\begin{equation}\label{eq:phonon_dispersion_q2}
\omega(q)=\frac{\sqrt{\mu(\lambda+2\mu)}}{\Gamma}\,q^2~.
\end{equation}
By contrast, long-range Coulomb interactions qualitatively modify the energetics of compressional deformations. Because the Coulomb kernel is nonlocal, the longitudinal elastic modulus becomes nonanalytic at small momentum, and the resulting low-energy collective mode crosses over to the characteristic magnetophonon scaling
\begin{equation}
\omega(q)\sim  q^{3/2}~, \qquad q\to 0~,
\end{equation}
as in the Tkachenko-type mode of crystals~\cite{PhysRevResearch.6.L012040,PhysRevB.109.035135,PhysRevB.110.035164,10.21468/SciPostPhys.17.6.164}. More generally, the universal Berry-phase term pairs the two displacement components into a single type-B Goldstone mode. For local short-range elasticity one finds a quadratic dispersion $\omega(q)\sim q^2$, whereas in the presence of unscreened long-range Coulomb interactions the nonlocal longitudinal stiffness produces the asymptotic $\omega(q)\sim q^{3/2}$ behavior, see Fig.~\ref{dispersion}.

A weak harmonic pinning potential, e.g.\ $\mathcal E_{\rm pin}=\frac12\kappa_p \bm u^2$, gaps the mode and produces a pinned resonance at
\begin{equation}
\omega_{\rm p}\sim \kappa_p/\Gamma~,
\end{equation}
which yields a sharp peak in the optical conductivity $\sigma_{xx}(\omega)$ and provides an experimentally accessible signature of skyrmion-crystal order.

\paragraph*{Including long-range interactions.} Long-range forces enter through density modulations induced by compressional deformations, $\delta\rho_{\rm el}= -\kappa e \bar\rho_{\rm sk}\,\nabla \cdot \bm u+\cdots$. Coupling $\delta\rho_{\rm el}$ to an interaction kernel $V_C(\bm q)$ produces
\begin{equation}\label{eq:int_u}
S_{\rm int}[\bm u]=-\frac12\int dt\sum_{\bm q}\;V_C(\bm q)\,|\delta\rho_{\rm el}(\bm q,t)|^2~.
\end{equation}
With this term included, the long-wavelength effective field theory is given by
\begin{equation}\label{eq:final_skyrmion_crystal_eft}
S_{\rm SkX}[\bm u;A] =S_{\rm B}[\bm u]+S_{\rm el}[\bm u]+S_{\rm int}[\bm u]
+\frac{\kappa e}{2\pi}\int d^3x\;\varepsilon^{\mu\nu\rho}A_\mu\,\partial_\nu a_\rho\!\left[\hat{\bm n}_0(\bm r-\bm u)\right]+\cdots~,
\end{equation}
which captures the universal Berry-phase kinematics, elastic energetics, Coulomb-modified magnetophonons, and texture-induced electromagnetic pumping in a unified framework.

\section{Noncommutative Phonons and Magnetic Translations of a Skyrmion Crystal}
\label{sec:noncommutative_magnetic_translations}

Throughout this section we assume the gapped Hall condition invoked in Sec.~\ref{sec:effective_field_theory}, so that the universal Berry term~\eqref{eq:berry_u} governs the long-wavelength kinematics. This first-order term endows the displacement field with a symplectic structure in which $u_x$ and $u_y$ are canonically conjugate, implying noncommutative phonon coordinates at long wavelengths. We now develop the associated centrally extended translation algebra and its magnetic-translation-group structure, and we connect it to a $W_\infty$/area-preserving-diffeomorphism kinematic organization.

\subsection{Central extension of translations}
For a crystal, continuous translations act on the phonon field as a uniform shift 
\begin{equation}\label{eq:shift_symmetry}
u_i(\bm r,t)\to u_i(\bm r,t)+a_i~,    
\end{equation}
when pinning is neglected. The Berry term is invariant under~\eqref{eq:shift_symmetry} up to a total time derivative:
\begin{equation}
\delta \mathcal L_B =\frac{\Gamma}{2}\varepsilon^{ij}a_i \partial_t u_j =\partial_t \left(\frac{\Gamma}{2}\varepsilon^{ij}a_i u_j\right)~.
\end{equation}
The corresponding choice of Noether generators is
\begin{equation}\label{eq:Pi_def}
P_i \equiv \Gamma\int d^2r\;\varepsilon_{ij}u_j(\bm r)~, \qquad \Gamma=\kappa\bar\rho_{\rm sk}~,
\end{equation}
which satisfies $i[P_i,u_j(\bm r)]=\delta_{ij}$. Using~\eqref{eq:noncomm_u_polished} gives the central extension
\begin{equation}\label{eq:Pxy_central_extension}
[P_x,P_y]
=i \Gamma\int d^2r
=i \Gamma A
=i \kappa \bar\rho_{\rm sk} A
=i \kappa Q^{\rm tot}~,
\end{equation}
where $A$ is the sample area and $Q^{\rm tot}=\bar\rho_{\rm sk} A$ is the total skyrmion number. The commutator density is finite and universal,
\begin{equation}
-\frac{i}{A}[P_x,P_y]=\Gamma=\kappa\bar\rho_{\rm sk}~.
\end{equation}
By the Watanabe-Murayama counting rule~\cite{PhysRevLett.110.181601,Watanabe:2019xul}, the nonzero rank of this commutator implies that the two broken translations pair into a single type-B Nambu-Goldstone mode in the clean, unpinned limit~\footnote{Let $\{Q_a\}$ denote the conserved charges associated with continuous symmetries of the microscopic Hamiltonian, and let $N_{\rm BS}$ be the number of spontaneously broken generators in the ground state. For nonrelativistic systems, the number $N_{\rm NG}$ of Nambu-Goldstone (NG) modes satisfies the Watanabe-Murayama counting rule,
\begin{equation}\label{eq:WM_counting}
N_{\rm NG} = N_{\rm BS} -\frac{1}{2}\mathrm{rank}\, \rho~, \qquad \rho_{ab}\equiv -\frac{i}{A} \big\langle [Q_a,Q_b]\big\rangle~,
\end{equation}
where $A$ is the system area and $\langle\cdots\rangle$ is the ground-state expectation value. When $\rho\neq 0$, pairs of broken generators become canonically conjugate and correspond to type-B NG modes typically with quadratic dispersion in $q$. When $\rho=0$, each broken generator yields an independent type-A NG mode (often with linear dispersion).

For a clean, unpinned skyrmion crystal in 2D, the broken continuous symmetries always include translations $T_x,T_y$, equivalently momentum operators $P_x,P_y$, hence $N_{\rm BS}\ge 2$. Whether the skyrmion crystal has one or two gapless phonons is determined by $\rho_{xy}$.

Plugging Eq.~\eqref{eq:Pxy_central_extension} into the counting formula~\eqref{eq:WM_counting} for the two broken translation generators gives
\begin{equation}\label{eq:counting_result}
N_{\rm BS}=2~, \qquad \rho_{xy}=-\rho_{yx}=\Gamma\neq 0 \Rightarrow \mathrm{rank} \, \rho=2 \Rightarrow N_{\rm NG}=2-\frac{1}{2}\times 2 = 1~.
\end{equation}
Therefore a 2D skyrmion crystal in a gapped quantum Hall background has one gapless phonon: a type-B NG mode. This should be contrasted with an ordinary non-magnetic crystal, where $\Gamma=0$ and one obtains two type-A acoustic phonons.}.

\subsection{Magnetic translation group}
Define finite translation operators acting on the skyrmion crystal collective coordinates by
\begin{equation}
T(\bm a)\equiv e^{\,i a_i P_i}~.
\end{equation}
Since the commutator~\eqref{eq:Pxy_central_extension} is a $c$-number, the Baker-Campbell-Hausdorff formula closes and yields
\begin{equation}\label{eq:magnetic_translation_group_law}
T(\bm a)\,T(\bm b)
=\exp\left[-\frac{1}{2}a_i b_j [P_i,P_j]\right]T(\bm a+\bm b)
=\exp\left[-\frac{i}{2}\Gamma A\,(\bm a\wedge\bm b)\right]T(\bm a+\bm b)~,
\end{equation}
equivalently,
\begin{equation}\label{eq:magnetic_translation_commutator}
T(\bm a)\,T(\bm b) =e^{-i\Gamma A(\bm a\wedge\bm b)}\,T(\bm b)\,T(\bm a)~.
\end{equation}
Thus skyrmion-crystal translations realize a projective magnetic representation with an effective ``flux density'' set by $\Gamma=\kappa\bar\rho_{\rm sk}$.

\paragraph*{Moir\'e pinning.} Explicit moir\'e pinning breaks continuous translations to a discrete subgroup and gaps the magnetophonon into a pinned resonance (pseudo-Goldstone).  Nevertheless, for weak pinning the local symplectic structure~\eqref{eq:noncomm_u_polished} remains the appropriate low-energy kinematics and controls, e.g., oscillator strengths and selection rules in optical response.

\subsection{Noncommutative phonons and \texorpdfstring{$W_\infty$}{W∞}/APD kinematics}

In the incompressible limit, the canonical structure promotes area-preserving diffeomorphisms (APDs)~\cite{10.21468/SciPostPhys.12.2.050,PhysRevResearch.4.033131,PhysRevResearch.6.L012040,PhysRevB.109.035135,du2025chiralgravitontheoryfractional} as an organizing principle for skyrmion crystal hydrodynamics, in close analogy to the APD structure of Landau-level dynamics. It is useful to distinguish the kinematic consequences of Eq.~\eqref{eq:noncomm_u_polished} from exact symmetries. In the approximately incompressible sector,
\begin{equation}\label{eq:incompressible_u}
\nabla\cdot\bm u\simeq 0~,\qquad u_i(\bm r,t)=\ell_u^2\varepsilon_{ij}\partial_j\phi(\bm r,t)~,
\end{equation}
deformations are divergence-free and can be parameterized by a scalar field $\phi$. Classically, divergence-free vector fields form the Lie algebra of area-preserving diffeomorphisms:
for $\xi_i(\lambda)\equiv \ell_u^2 \varepsilon_{ij}\partial_j\lambda$,
\begin{equation}\label{eq:APD_algebra}
[\xi(\lambda_1),\xi(\lambda_2)]_i=\xi_i(\{\lambda_1,\lambda_2\})~,
\qquad \{\lambda_1,\lambda_2\}\equiv \ell_u^2\varepsilon^{ij}\partial_i\lambda_1 \partial_j\lambda_2~.
\end{equation}
Quantization of the noncommutative phonon phase space then provides a long-wavelength realization of a $W_\infty$ algebra, directly analogous to the guiding-center $W_\infty$/GMP structure in quantum Hall fluids. Crucially, $W_\infty$/APD should be viewed here as an organizing kinematic structure inherited from the universal Berry phase term, rather than an exact symmetry of the full skyrmion-crystal action: elastic energetics and moir\'e pinning select a background lattice/metric and reduce APD down to the appropriate space-group and, with pinning, to a discrete subgroup.

\paragraph*{GMP algebra for skyrmion density.} A compact way to expose the GMP structure is to express the skyrmion density of a deformed configuration using the displaced coordinates $\bm X(\bm r,t)\equiv \bm r-\bm u(\bm r,t)$.
At the nonlinear level one may write
\begin{equation}\label{eq:rho_pullback}
\rho(\bm q,t)\equiv \int d^2r\;\rho_0(\bm r)\,e^{-i\bm q\cdot\bm X(\bm r,t)}~,
\end{equation}
with $\rho_0(\bm r)$ the static skyrmion density of $\hat{\bm n}_0$. Using Eq.~\eqref{eq:noncomm_u_polished}, smooth long-wavelength deformations imply a closed density algebra of the form
\begin{equation}\label{eq:SkX_GMP_like}
[\rho(\bm q),\rho(\bm k)]=2i\sin\left(\frac{\ell_u^2}{2}\bm q\wedge \bm k\right)\rho(\bm q+\bm k)+\cdots~,
\end{equation}
where the ellipsis denotes higher-gradient corrections and lattice-anisotropy effects set by $\mathcal E_{\rm el}$ and $\mathcal E_{\rm pin}$. These formalisms make explicit that the skyrmion-crystal phonons realize a noncommutative kinematics, while the parity-even energetics determine the resulting magnetophonon dynamics.

\subsection{Noncommutative field theory and decay rate}

The universal Berry phase term in the magneto-elastic action, $\mathcal L_B=\frac{\Gamma}{2}\varepsilon_{ij}u_i\partial_t u_j$ with $\Gamma=\kappa\bar\rho_{\rm sk}$, endows the skyrmion-crystal phonons with a first-order symplectic structure and hence noncommutative kinematics. At long wavelengths and for weak pinning this places the magnetophonon in the same noncommutative effective field theory (NCFT) class~\cite{Rubakov-NC,RevModPhys.73.977,Du_2024} as the Tkachenko mode of a vortex lattice~\cite{PhysRevResearch.6.L012040,PhysRevB.109.035135,PhysRevB.110.035164}. Besides providing a compact single-field description, the NCFT viewpoint is useful because it organizes nonlinearities and constrains damping processes by magnetic-translation symmetry.

We define the long-wavelength noncommutativity parameter
\begin{equation}\label{eq:theta_u_def}
\theta_u \equiv \ell_u^2=\frac{1}{\Gamma}=\frac{1}{\kappa\bar\rho_{\rm sk}}~,
\end{equation}
so that, locally, $u_x$ and $u_y$ form a canonically conjugate pair
\begin{equation}
[u_x(\bm r),u_y(\bm r')]=i\theta_u \delta(\bm r-\bm r')+\cdots~.    
\end{equation}

\paragraph*{Moyal product and noncommutative coordinates.} Introduce the Moyal star product,
\begin{equation}\label{eq:star_product_u}
(f\star g)(\bm r)\equiv f(\bm r)\,\exp\!\left(-\frac{i}{2}\theta_u\,\varepsilon^{ij}\overleftarrow{\partial_i}\overrightarrow{\partial_j}\right)\,g(\bm r)~, \qquad [f,g]_\star \equiv f\star g-g\star f~.
\end{equation}
Define the material coordinates of the crystal,
\begin{equation}\label{eq:material_coords}
X^a(\bm r,t)\equiv r^a-u^a(\bm r,t)~,\qquad a=x,y~.
\end{equation}
In the incompressible/area-preserving regime, equivalently, after integrating out the high-energy compressional response at long wavelength, the nonlinear constraint can be implemented in NCFT as
\begin{equation}\label{eq:XY_constraint}
[X^1,X^2]_\star = -i\theta_u~,
\end{equation}
which admits a general solution in terms of a single real scalar field $\phi(\bm r,t)$,
\begin{equation}\label{eq:X_unitary_phi}
X^a(\bm r,t)=U_\phi(\bm r,t)\star r^a \star U_\phi(\bm r,t)^{-1}~, \qquad U_\phi \equiv e_\star^{\,i\phi}=1+i\phi - \frac{1}{2} \phi \star \phi +\cdots~.
\end{equation}
Expanding \eqref{eq:X_unitary_phi} to leading order gives the divergence-free form
\begin{equation}\label{eq:u_phi_linear}
u^a(\bm r,t)=r^a-X^a(\bm r,t) = \theta_u \varepsilon^{ab}\partial_b\phi(\bm r,t)+O(\theta_u^3\partial^3\phi)~,
\end{equation}
consistent with Eq.~\eqref{eq:incompressible_u}~\footnote{Define star-covariant objects
\begin{equation}\label{eq:Dmu_phi_def}
D_\mu\phi \equiv -i(\partial_\mu U_\phi)\star U_\phi^{-1},\qquad \mu\in\{t,x,y\}~,
\end{equation}
so that $D_\mu\phi=\partial_\mu\phi+O(\theta_u)$. Then~\eqref{eq:X_unitary_phi} can be written to all orders as
\begin{equation}\label{eq:X_all_orders}
X^a = r^a + \theta_u \varepsilon^{ai} D_i\phi~,
\end{equation}
and the most general local Lagrangian consistent with magnetic translations can be organized as
\begin{equation}\label{eq:L_general_D0_Dab}
\mathcal L_\phi = \mathcal L\!\left(D_0\phi,\; D_{ab}\phi,\;\ldots\right),
\end{equation}
where $D_{ab}\phi$ is the symmetric traceless combination built from spatial covariant derivatives (as in Ref.~\cite{PhysRevResearch.6.L012040}).}.

\paragraph*{Quadratic Lifshitz form and coefficient matching.} Magnetic-translation invariance restricts the quadratic $\phi$ theory to a Lifshitz form,
\begin{equation}\label{eq:L_phi2_Lifshitz}
\mathcal L_{\phi,2} =\frac{c_0}{2}(\partial_t\phi)^2-\frac{c_1}{2}(\nabla^2\phi)^2+\cdots~,
\end{equation}
as in Ref.~\cite{PhysRevResearch.6.L012040}. Matching~\eqref{eq:L_phi2_Lifshitz} to the isotropic magneto-elastic theory~\eqref{eq:SkX_EFT_final} gives
\begin{equation}\label{eq:c0c1_matching}
c_0=\frac{1}{\lambda+2\mu}~,\qquad c_1=\frac{\mu}{\Gamma^2}~,\qquad \Gamma=\kappa \bar\rho_{\rm sk}~,
\end{equation}
and hence $\omega(q)=\sqrt{\mu(\lambda+2\mu)}\,q^2/\Gamma$ in the short-range-elasticity regime, consistent with Eq.~\eqref{eq:phonon_dispersion_q2}. Long-range Coulomb interactions enter as a nonlocal enhancement of the longitudinal modulus, producing a nonlocal modification of~\eqref{eq:L_phi2_Lifshitz} and the crossover to $\omega\sim q^{3/2}$ at small $q$.

\paragraph*{Magnetic translations as noncommutative dipole symmetry.} Because translations are centrally extended in the skyrmion-crystal phase space, they act on $U_\phi$ as a special class of star-unitary transformations generated by linear functions of $\bm r$, implying a noncommutative (magnetic) generalization of dipole symmetry in the $\phi$ description~\cite{PhysRevResearch.6.L012040}.

\paragraph*{Intrinsic decay rate in the clean limit.} In the clean, unpinned regime with short-range elasticity where $\omega\sim q^{2}$,
a $1 \to 2$ decay of one magnetophonon into two lower-energy magnetophonons is kinematically allowed. Magnetic-translation invariance strongly constrains the leading cubic interactions in the NCFT; by direct analogy with the Tkachenko-mode analysis of Ref.~\cite{PhysRevResearch.6.L012040}, this implies that at zero temperature $T=0$ the intrinsic decay rate obeys the low-energy scaling
\begin{equation}
\gamma(E)\sim E^{3}~,\qquad \omega \sim q^2~,
\end{equation}
so that $\gamma(E)/E\to 0$ as $E\to 0$ and the magnetophonon becomes asymptotically sharp. This scaling is expected to cross over when long-range Coulomb interactions control the asymptotic dispersion, and in realistic moir\'e samples it can be masked by extrinsic disorder/pinning which often dominates the linewidth at the lowest frequencies.

\section{Conclusion and Outlook}
\label{sec:conclusion}

We presented a controlled and largely analytical framework for electrons coupled to smooth moir\'e-periodic skyrmion textures, aimed at connecting real-space texture geometry to quantum geometry diagnostics and collective dynamics in moir\'e Chern-band settings. The central step is an exact local $SU(2)$ rotation into the texture-aligned frame, which generates an emergent non-Abelian gauge structure from spatial variation of the instantaneous pseudospin eigenbasis. In the regime where two local branches are separated by a large splitting $2J(\bm r)$ and the texture varies slowly, we performed an operator-level Schrieffer-Wolff expansion in $1/J$ that integrates out the high-energy branch.

A key conceptual point for the many-body interpretation is that the SW procedure yields not only an effective single-branch Hamiltonian but also a systematic mapping of microscopic operators into the low-energy subspace. The physical density, current, and other observables are dressed by the same unitary that block-diagonalizes the Hamiltonian. Together, the effective Hamiltonian and dressed operators define the projected interacting theory: interactions retain their standard density-density form, but act through a dressed projected density operator and its associated form factors. This operator-level control provides a microscopic starting point for response and effective-field-theory constructions without assuming ideal Landau-level kinematics.

On the single-branch side, we reorganized the dynamics into a uniform-plus-periodic Landau-level representation: motion in the spatially averaged emergent field $b_0$ is perturbed by moir\'e-periodic magnetic and scalar modulations, together with controlled $O(J^{-1})$ corrections fixed by the texture's real-space quantum geometry. In the long-wavelength/LLL regime, flux inhomogeneity produces a smooth deformation of guiding-center noncommutativity and hence an Umklapp-resolved deformation of GMP kinematics. This provides a concrete route to quantify departures from ideal Landau-level behavior directly in terms of texture data (emergent flux and real-space quantum geometric tensor).

Building on this structure, we connected an experimentally accessible many-body diagnostic, the long-wavelength longitudinal optical quantum weight obtained from an optical sum rule, to the general topological lower bound for gapped Hall phases~\cite{PhysRevX.14.011052}. Within the present microscopic framework we identified two controlled sources of excess weight above the topological minimum: Berry-curvature harmonics associated with flux inhomogeneity and moir\'e periodicity, and finite-$J$ inter-branch mixing encoded by the leading SW correction and the accompanying operator dressing.

To access experimentally relevant spectra in the presence of moir\'e periodicity, we formulated Umklapp-resolved density form factors in a magnetic Bloch basis and developed a matrix response formalism. Within a standard TDHF/RPA approximation that retains Umklapp mixing explicitly, we showed that mode folding is generic: collective modes residing primarily at finite Umklapp momentum can acquire oscillator strength in the nominally uniform channel. This provides a direct mechanism for additional optically active neutral modes in long-wavelength probes even when the underlying excitation ``lives'' at finite crystal momentum.

Finally, under a clearly stated many-body assumption, we derived an effective field theory for a skyrmion crystal. Assuming the projected interacting system realizes a gapped Hall phase with coefficient $\kappa$, integrating out the gapped electronic sector yields a Chern-Simons term for the combined gauge field $B_\mu\equiv eA_\mu+a_\mu[\hat{\bm n}]$. Expanding about a periodic skyrmion-crystal saddle point then produces a magneto-elastic theory for deformations with a universal parity-odd Berry term fixed by $\kappa$ and the mean skyrmion density. This term renders $u_x$ and $u_y$ canonically conjugate, implying noncommutative phonon coordinates and pairing the two broken translations into a single type-B magnetophonon. For short-range elasticity, the magnetophonon dispersion is given by $\omega\sim q^2$; with long-range Coulomb interactions the mode crosses over to $\omega\sim q^{3/2}$; weak pinning yields a pinned resonance.

\paragraph*{Experimental outlook.} Our results suggest several experimentally accessible signatures in twisted TMD homobilayers and rhombohedral graphene aligned with hBN. First, broadband THz/infrared measurements of long-wavelength optical conductivity $\mathrm{Re}\,\sigma_{xx}(\omega)$ can access the long-wavelength optical quantum weight through the sum rule, providing a direct probe of departures from ideal Landau-level geometry. Within our framework, tuning parameters that control flux inhomogeneity and the effective splitting $J$ offers a route to disentangle the two controlled microscopic sources of excess weight. Second, Umklapp-induced mode folding implies that additional neutral collective modes can become optically active in the nominally long-wavelength channel, especially relevant for THz and Raman probes. Third, in regimes with skyrmion-crystal order on a gapped Hall background, weak pinning (from moir\'e commensurability or disorder) gaps the magnetophonon into a pinned resonance, yielding a characteristic low-frequency peak in $\sigma_{xx}(\omega)$ and a direct signature of crystalline order and emergent-flux dynamics at zero applied field. It would be particularly valuable to design momentum-resolved experiments that directly map the magnetophonon dispersion and its crossover behavior; we defer a concrete experimental proposal to future work.

\paragraph*{Theoretical outlook.} Several extensions follow naturally. A first direction is quantitative microscopic matching: extracting parity-even coefficients entering the skyrmion-crystal EFT, e.g. elastic moduli, pinning scales, and controlled $1/J$ renormalizations, directly from continuum moir\'e models or Hartree-Fock energy functionals, enabling parameter-controlled predictions for collective-mode frequencies and optical spectral weights. A second direction is to incorporate disorder, dissipation, and finite temperature into the Umklapp-resolved response and into the magneto-elastic theory, in order to model linewidths and depinning and connect more directly to experimental lineshapes. A third direction is to generalize the operator-level projection to multi-component manifolds, e.g., multiple valleys/spins or nearly degenerate bands, where the projected Berry connection and quantum geometry become genuinely non-Abelian and additional internal collective modes and geometric responses~\cite{WenZeePRL1992Shift,HoyosSon2012HallViscosity,Son2013NewtonCartan} may emerge. It would also be interesting to extend the analysis developed here systematically to rhombohedral $N$-layer graphene aligned with hBN (RNG/hBN), and to work out the corresponding microscopic matching and collective-mode phenomenology; we defer this to future work. More broadly, the controlled projection framework developed here provides a systematic route for studying how real-space texture geometry, flux inhomogeneity, and finite-$J$ mixing renormalize operator algebra and measurable response in moir\'e Chern-band settings while keeping moir\'e periodicity explicit.

\paragraph*{Note added.}
While this paper was being completed, we became aware of Ref.~\cite{tan2025ideallimitrhombohedralgraphene}, which has partial overlap with our work. Our emphasis differs in developing an operator-level controlled $1/J$ expansion beyond strict adiabaticity for generic texture-induced emergent fields, formulating Umklapp-resolved response and mode folding, and deriving a skyrmion-crystal magneto-elastic effective theory whose parity-odd structure is fixed by quantized electronic response.

\acknowledgments
We thank Ahmed Abouelkomsan, Liang Fu, Nuh Gedik, Tonghang Han, Jung Hoon Han, Long Ju, Patrick A. Lee, Leonid Levitov, Nicolás Morales-Durán, Nisarga Paul, Inti Sodemann, Dam Thanh Son, Alexander von Hoegen, Xiao-Gang Wen, and Xiaodong Xu for related discussions. This work is supported, in part by the Simons Collaboration on Ultra-Quantum Matter, which is a grant from the Simons Foundation (651446). This work was performed in part at the Aspen Center for Physics, which is supported by a grant from the Simons Foundation (1161654, Troyer). 

\appendix

\section{Schrieffer-Wolff generator at leading order}
\label{app:SW_generator}

In this Appendix, we give an equivalent expression that follows from the standard Schrieffer–Wolff generator construction. Split $\tilde H = H_0+V$ with $H_0=-J(\bm r)\sigma_z$ and $V$ the remaining terms, and decompose $V=V_{\rm d}+V_{\rm od}$ into diagonal/off-diagonal parts in the $\sigma_z$ basis. Choose an anti-Hermitian off-diagonal generator $S=-S^\dagger$ such that $[H_0,S]=V_{\rm od}$, where $V_{\rm od}=H_{+-}\sigma_+ + H_{-+}\sigma_-$. To leading order in $1/J$ this gives
\begin{equation}
S = \frac{1}{2J(\bm r)}
\begin{pmatrix}
0 &- H_{+-}\\
 H_{-+} & 0
\end{pmatrix}
+O(J^{-2})~.
\end{equation}

Projecting onto the low-energy branch with $P_+=(1+\sigma_z)/2$, the SW expansion yields
\begin{equation}
H_{\rm eff}=P_+\Big(H_0+V_{\rm d}+\frac12[S,V_{\rm od}]\Big)P_+ + O(J^{-2})~,
\end{equation}
and therefore, to leading nontrivial order gives Eq.~\eqref{eq:HeffSWfinal}.

\section{Non-Abelian Gauge-Field Identities and Projection}
\label{app:nonabelian}

In this Appendix, we discuss the exact non-Abelian gauge structure generated by the local $SU(2)$ rotation into the texture-aligned frame and derive identities that connect the projected $U(1)$ emergent field and real-space quantum geometry to the off-diagonal non-adiabatic gauge couplings.

\paragraph*{Pure-gauge $\mathfrak{su}(2)$ connection and vanishing curvature.} Let $U(\bm r,t)\in SU(2)$ satisfy $U^\dagger(\bm r,t)\hat{\bm n}(\bm r,t) \cdot \bm\sigma U(\bm r,t)=\sigma_z$~. Define the emergent non-Abelian connection
\begin{equation}
\mathcal A_\mu(\bm r,t)\equiv iU^\dagger \partial_\mu U \in \mathfrak{su}(2)~,
\qquad \mu\in\{t,x,y\}~.
\end{equation}
Under an $SU(2)$ change of local frame $U\to U V$ one has the standard gauge transformation
\begin{equation}
\mathcal A_\mu \to V^\dagger \mathcal A_\mu V + iV^\dagger \partial_\mu V~.
\end{equation}
Because $\mathcal A_\mu$ is generated by a basis rotation, it is a pure gauge:
\begin{equation}\label{eq:nonabelian_curvature_zero}
\mathcal F_{\mu\nu}\equiv \partial_\mu\mathcal A_\nu-\partial_\nu\mathcal A_\mu-i[\mathcal A_\mu,\mathcal A_\nu]=0~.
\end{equation}
which holds in the full two-component Hilbert space and is independent of any adiabatic approximation.

\paragraph*{Residual $U(1)$ gauge freedom.} Imposing $U^\dagger \hat{\bm n}\cdot\bm\sigma\,U=\sigma_z$ leaves a residual
$U(1)$ gauge freedom $U\to U e^{i\varphi\sigma_z/2}$. In the $\sigma_z$ basis it is convenient to decompose
\begin{equation}\label{eq:A_mu_decomp_app}
\mathcal A_\mu=
\begin{pmatrix}
-a_\mu & W_\mu\\
W_\mu^\dagger & a_\mu
\end{pmatrix}~,
\qquad
a_\mu\equiv -(\mathcal A_\mu)_{++}~,\quad
W_\mu\equiv (\mathcal A_\mu)_{+-}~.
\end{equation}
Under the residual $U(1)$ transformation,
\begin{equation}\label{eq:U1_trans_app}
a_\mu \to a_\mu + \frac{1}{2}\partial_\mu\varphi~,
\qquad
W_\mu \to e^{-i\varphi} W_\mu~,
\end{equation}
so $W_\mu$ carries unit $U(1)$ charge and mediates inter-branch (non-adiabatic) mixing.

For the static texture parameterization
$\hat{\bm n}=(\sin\theta\cos\phi,\sin\theta\sin\phi,\cos\theta)$, one convenient gauge choice gives
\begin{align}
a_i &= -i\langle u_+|\partial_i u_+\rangle
=\frac{1-\cos\theta}{2}\partial_i\phi~,\\
W_i &= i\langle u_+|\partial_i u_-\rangle
=-\frac{e^{-i\phi}}{2}\big(\sin\theta\,\partial_i\phi+i\partial_i\theta\big)~,
\end{align}
consistent with Eqs.~\eqref{eq:ai_theta_phi}--\eqref{eq:Wi_theta_phi}.

\paragraph*{Exact identities from $\mathcal F_{\mu\nu}=0$.} Substituting~\eqref{eq:A_mu_decomp_app} into~\eqref{eq:nonabelian_curvature_zero} yields component-wise identities. For spatial indices $i,j\in\{x,y\}$,
\begin{equation}\label{eq:Fij_components}
0=\mathcal F_{ij}=
\begin{pmatrix}
-f_{ij}-i(W_iW_j^\dagger-W_jW_i^\dagger) & (D_i W_j-D_j W_i)\\
(D_i W_j-D_j W_i)^\dagger & -f_{ij}-i(W_i^\dagger W_j-W_j^\dagger W_i)
\end{pmatrix}~,
\end{equation}
where
\begin{equation}
f_{ij}\equiv \partial_i a_j-\partial_j a_i~,
\qquad
D_i W_j \equiv (\partial_i+2ia_i)W_j
\end{equation}
is the residual-$U(1)$ covariant derivative consistent with~\eqref{eq:U1_trans_app}.

Eq.~\eqref{eq:Fij_components} implies two exact relations:
\begin{align}
\partial_i a_j-\partial_j a_i
&= -i\big(W_iW_j^\dagger-W_jW_i^\dagger\big)~,
\label{eq:curl_a_W_exact}
\\
D_i W_j - D_j W_i &= 0~.
\label{eq:covcurl_W_exact}
\end{align}

\paragraph*{Projected $U(1)$ emergent field and quantum geometry.} The emergent projected $U(1)$ flux density in the adiabatic branch is
\begin{equation}
b(\bm r)\equiv \varepsilon^{ij}\partial_i a_j(\bm r)~.
\end{equation}
From~\eqref{eq:curl_a_W_exact},
\begin{equation}\label{eq:b_from_W}
b(\bm r)= -i\varepsilon^{ij}W_i(\bm r)W_j^\dagger(\bm r) =2\Im \big(W_x W_y^\dagger\big)~.
\end{equation}

Define the real-space quantum geometric tensor of the local adiabatic spinor $|u_+(\bm r)\rangle$,
\begin{equation}
\mathcal Q_{ij}(\bm r)\equiv \langle \partial_i u_+|(1-|u_+\rangle\langle u_+|)|\partial_j u_+\rangle \equiv g_{ij}+\frac{i}{2}\Omega_{ij}~.
\end{equation}
For a two-level system, $(1-|u_+\rangle\langle u_+|)=|u_-\rangle\langle u_-|$, and one finds the exact identity
\begin{equation}\label{eq:Tij_equals_WW_app}
\mathcal Q_{ij}(\bm r)=W_i(\bm r)\,W_j^\dagger(\bm r)~,
\end{equation}
so that
\begin{equation}
g_{ij}=\Re\,\mathcal Q_{ij},\qquad \Omega_{ij}=2\Im\,\mathcal Q_{ij}~.
\end{equation}
Combining with~\eqref{eq:b_from_W} gives the compact equality
\begin{equation}\label{eq:b_equals_Omega}
b(\bm r)=\Omega_{xy}(\bm r)
=\frac12 \hat{\bm n}\cdot(\partial_x\hat{\bm n}\times\partial_y\hat{\bm n})~,
\end{equation}
which is the microscopic bridge between the projected emergent flux density and real-space quantum geometry.

Likewise the geometric scalar potential is
\begin{equation}
\Phi_{\rm g}(\bm r)=\frac{1}{2m}\sum_i W_iW_i^\dagger =\frac{1}{2m}\Tr g(\bm r)=\frac{1}{8m}(\partial_i\hat{\bm n})^2~,
\end{equation}
reproducing Eq.~\eqref{eq:BO_potential} and making explicit that $\Phi_{\rm g}$ is fixed by the local quantum metric.

\paragraph*{Projection produces a nonzero ``magnetic field.''} The vanishing of the full curvature $\mathcal F_{ij}=0$ does not contradict the existence of a nonzero projected flux $b(\bm r)$. Eq.~\eqref{eq:curl_a_W_exact} shows that the curl of the diagonal component $a_i$ is cancelled in the full $SU(2)$ curvature by the commutator term involving the off-diagonal mixing fields $W_i$. Upon projecting to a single branch, this cancellation is no longer available, and $b(\bm r)$ becomes the physical Berry curvature experienced by the adiabatic electron.

\paragraph*{Exact (non-perturbative) reduction of the high-energy branch.} Keeping the full non-Abelian gauge field corresponds simply to working with the exact rotated-frame two-component Hamiltonian
\begin{equation}
\tilde H=\frac{(\bm p-\bm{\mathcal A})^2}{2m}-J(\bm r)\sigma_z+V(\bm r)~,
\end{equation}
without any adiabatic approximation. The effect of branch mixing can be retained non-perturbatively at the level of the low-energy
resolvent via the Schur complement (Feshbach) identity:
\begin{equation}\label{eq:Schur_exact_app}
H_{\rm eff}(E)=H_{++}-H_{+-}\,\frac{1}{H_{--}-E}\,H_{-+}~.
\end{equation}
Eq.~\eqref{eq:Schur_exact_app} is exact and keeps the non-Abelian mixing encoded in $W_i$ to all orders, but is energy dependent and generally nonlocal as an operator. Expanding $(H_{--}-E)^{-1}$ in $1/J$ reproduces the controlled SW series used in the main text.

In the present single-valley/two-level setting, the low-energy adiabatic subspace is one-dimensional, so the Berry connection after projection is necessarily $U(1)$ (Abelian). A genuine non-Abelian Berry connection and non-Abelian quantum geometry arise only when projecting onto an $N>1$ nearly degenerate manifold, e.g., multiple valleys/spins with symmetry-protected degeneracy, yielding a $U(N)$ connection with nontrivial curvature. Our $SU(2)$ connection $\mathcal A_\mu=iU^\dagger\,\partial_\mu U$ should be viewed instead as a pure-gauge connection on the full two-component Hilbert space whose off-diagonal components control non-adiabatic mixing.

\section{Systematic Mapping of the Controlled \texorpdfstring{$1/J$}{1/J} Correction to the Skyrmion-Crystal Effective Field Theory}
\label{sec:map_1overJ_to_EFT}

In this Appendix, we discuss how the controlled Schrieffer-Wolff correction at order $1/J$ maps into the long-wavelength effective field theory of a skyrmion crystal. The key point is that the SW procedure yields a well-defined single-branch electronic theory whose deviations from the strict adiabatic limit are organized in powers of $1/J$.
When the resulting interacting electron problem is in a gapped Hall phase,
integrating out the electrons produces a topological Chern-Simons term with quantized coefficient $\kappa$ and a parity-even local functional whose coefficients, admitting an expansion in $1/J$, determine the elastic moduli and higher-gradient terms of the skyrmion-crystal effective field theory.

\paragraph*{Single-branch action and the controlled $1/J$ correction.} After the local-frame rotation and SW reduction described in the main text, the low-energy fermion field $\chi$ (one pseudospin branch) couples to the external electromagnetic probe $A_\mu$ and to the texture-induced Berry connection $a_\mu[\hat{\bm n}]$ only through the combined gauge field
\begin{equation}
B_\mu \equiv eA_\mu + a_\mu[\hat{\bm n}]~.
\end{equation}
Working in Euclidean spacetime $x\equiv(\tau,\bm r)$, we use the shorthand $\int d^3x\equiv \int_0^\beta d\tau\int d^2r$, where $\tau\in[0,\beta]$ is the imaginary-time coordinate and $\beta\equiv (k_B T)^{-1}$. Throughout, we set $k_B=1$. With this notation, a convenient form of the single-branch action is
\begin{equation}\label{eq:app_single_branch_action}
S_{\rm el}[\chi;B,\hat{\bm n}]
=\int d^3x\; \chi^\dagger\!\left[D_\tau^{(B)} + \mathcal H_{\rm ad}[\hat{\bm n}] + \delta \mathcal H_{1/J}[\hat{\bm n}]+ \mathcal O(J^{-2})
\right]\!\chi +S_{{\rm int},E}[\chi]~,
\end{equation}
where $D_\tau^{(B)}\equiv \partial_\tau-iB_\tau$ and
\begin{equation}
\mathcal H_{\rm ad}[\hat{\bm n}] =\frac{\big(\bm p-\bm B\big)^2}{2m}+V(\bm r)+\Phi_{\rm g}(\bm r)+\cdots~, \qquad \Phi_{\rm g}(\bm r)=\frac{1}{8m}(\partial_i\hat{\bm n})^2~.
\end{equation}
The leading non-adiabatic correction is the gauge-covariant SW term
\begin{equation}\label{eq:app_deltaH_SW}
\delta \mathcal H_{1/J} = -\mathcal H_{+-}\,\frac{1}{2J(\bm r)}\,\mathcal H_{-+} + \mathcal O(J^{-2})~,
\end{equation}
written in the texture-aligned frame, where $\mathcal H_{\pm\mp}$ are the off-diagonal operators that mix the two local pseudospin branches.

In the long-wavelength regime where $J(\bm r)$ and the texture vary slowly on the moir\'e scale, $\delta\mathcal H_{1/J}$ can be expressed in terms of the real-space quantum geometric tensor of the local adiabatic spinor. A compact leading-gradient form is
\begin{equation}\label{eq:app_deltaH_QGT}
\delta \mathcal H_{1/J} \simeq -\frac{1}{2m^2J(\bm r)} \pi_i
\left[g_{ij}(\bm r)+\frac{i}{2}\Omega_{ij}(\bm r)\right]\pi_j + \cdots~,
\end{equation}
where $\pi_i\equiv -i\partial_i-a_i$ is the branch-covariant momentum and $g_{ij}$ and $\Omega_{ij}$ are the real-space quantum metric and Berry curvature of the local adiabatic state. The ellipsis denotes higher-gradient and operator-ordering corrections, e.g., the Hermitian
symmetrization required when $J(\bm r)$ and $g_{ij}(\bm r)$ do not commute with gradients.

\paragraph*{Integrating out a gapped Hall sector: topological versus parity-even terms.} Assume that the interacting single-branch problem~\eqref{eq:app_single_branch_action} realizes a fully gapped quantum Hall phase. Integrating out $\chi$ defines an effective action for the slow collective fields $(A_\mu,\hat{\bm n})$,
\begin{equation}\label{eq:app_Seff_def}
e^{-S_{\rm eff}[A,\hat{\bm n}]} \equiv \int \mathcal D\chi^\dagger\mathcal D\chi\; e^{-S_{\rm el}[\chi;B,\hat{\bm n}]}~, \qquad
S_{\rm eff}=S_{\rm top}[B]+S_{\rm even}[B,\hat{\bm n}]+\cdots~.
\end{equation}
Gauge invariance and locality imply that the leading parity-odd term is the Chern-Simons response
\begin{equation}\label{eq:CS_top_term}
S_{\rm top}[B] = \frac{\kappa}{4\pi}\int d^3x\; \varepsilon^{\mu\nu\rho}B_\mu\partial_\nu B_\rho~,
\end{equation}
where $\kappa=(2\pi/e^2)\sigma_{xy}$ is the dimensionless Hall coefficient of the gapped many-body phase. Crucially, as long as the bulk gap and $U(1)$ charge conservation are preserved, $\kappa$ is quantized
and cannot be renormalized by smooth $1/J$ corrections. Therefore, the controlled non-adiabatic correction $\delta\mathcal H_{1/J}$ enters $S_{\rm eff}$ only through the parity-even functional $S_{\rm even}$ and higher-derivative terms.

\paragraph*{Cumulant expansion in $1/J$.} To make the $1/J$ renormalization of parity-even effective field theory coefficients explicit, split
\begin{equation}
S_{\rm el} = S_{0} + \delta S_{1/J}~, \qquad S_{0}\equiv S_{\rm el}\big|_{\delta\mathcal H_{1/J}=0},\qquad \delta S_{1/J}\equiv \int d^3x\;\chi^\dagger \delta\mathcal H_{1/J}\chi~.
\end{equation}
Define expectation values in the adiabatic theory by
\begin{equation}\label{eq:app_expectation_def}
\langle \mathcal O\rangle_{0} \equiv \frac{1}{Z_{0}} \int \mathcal D\chi^\dagger\mathcal D\chi\;\mathcal O\;e^{-S_{0}}~, \qquad
Z_{0}\equiv \int \mathcal D\chi^\dagger\mathcal D\chi\;e^{-S_{0}}~.
\end{equation}
Then
\begin{equation}
Z \equiv e^{-S_{\rm eff}} = Z_0\,\big\langle e^{-\delta S_{1/J}}\big\rangle_{0} \qquad \Rightarrow \qquad S_{\rm eff}=S_{\rm eff}^{(0)}-\ln\big\langle e^{-\delta S_{1/J}}\big\rangle_{0}~.
\end{equation}
Expanding the logarithm generates the standard cumulant series,
\begin{equation}\label{eq:app_cumulant}
\delta S_{\rm even}^{(1/J)} = \big\langle \delta S_{1/J}\big\rangle_{0}
-\frac{1}{2}\big\langle (\delta S_{1/J})^2\big\rangle^{c}_{0} +\cdots~,
\end{equation}
where the superscript $c$ denotes the connected correlator, e.g.
$\langle X^2\rangle_0^{c}\equiv \langle X^2\rangle_0-\langle X\rangle_0^2$. Equivalently, writing $\delta S_{1/J}=\int_0^\beta d\tau\,\delta H_{1/J}(\tau)$,
\begin{equation}\label{eq:app_cumulant_time}
\delta S_{\rm even}^{(1/J)} = \int_0^\beta d\tau\,\big\langle \delta H_{1/J}(\tau)\big\rangle_{0} -\frac{1}{2}\int_0^\beta d\tau\int_0^\beta d\tau'\,\big\langle T_\tau\,\delta H_{1/J}(\tau)\delta H_{1/J}(\tau')\big\rangle^{c}_{0} +\cdots~.
\end{equation}

For a Gaussian (mean-field) electronic theory, the same expansion may be written in trace form as
\begin{equation}\label{eq:app_trace_log}
\delta S_{\rm even}^{(1/J)} = -\Tr\left(G_{\rm ad}\,\delta\mathcal H_{1/J}\right) +\frac{1}{2}\Tr\left(G_{\rm ad}\,\delta\mathcal H_{1/J}\,G_{\rm ad}\,\delta\mathcal H_{1/J}\right)_{\!c} +\cdots~,
\end{equation}
where $G_{\rm ad}$ is the time-ordered Green's function of the adiabatic single-branch theory and $\Tr$ denotes a trace over spacetime as well as internal indices. In an interacting gapped phase,~\eqref{eq:app_cumulant} and \eqref{eq:app_cumulant_time} provide the most general definition;~\eqref{eq:app_trace_log} should then be viewed as shorthand for the corresponding connected current/energy correlators or as the appropriate approximation within the chosen many-body scheme.

\paragraph*{Specialization to a skyrmion crystal: universal Berry phase term.} Assume that the parity-even functional admits a static periodic saddle point $\hat{\bm n}_0(\bm r)$ carrying skyrmion number $Q$ per moir\'e unit cell of area $A_M$. Long-wavelength distortions are parameterized by a displacement field $\bm u(\bm r,t)$ via
\begin{equation}\label{eq:app_texture_displacement}
\hat{\bm n}(\bm r,t)=\hat{\bm n}_0(\bm r-\bm u(\bm r,t))+\cdots~, \qquad a_\mu(\bm r,t)=a_{\mu,0}(\bm r-\bm u(\bm r,t))+\cdots~.
\end{equation}
Substituting~\eqref{eq:app_texture_displacement} into the Chern-Simons term~\eqref{eq:CS_top_term} and expanding to leading order in $\partial_t\bm u$ yields the universal Berry-phase term for phonons,
\begin{equation}\label{eq:app_SB_u}
S_{B}[\bm u] = \frac{\Gamma}{2}\int dt\,d^2r\;\varepsilon_{ij} u_i\,\partial_t u_j+\cdots~, \qquad \Gamma=\kappa\bar\rho_{\rm sk}=\kappa \frac{Q}{A_M}~.
\end{equation}
This term pairs the two broken translations into a single type-B Goldstone mode and implies noncommutative phonon coordinates at long wavelengths. Importantly, since $\Gamma$ is fixed by the quantized Hall response $\kappa$ and the mean skyrmion density $Q/A_M$, it is not renormalized by smooth $1/J$ corrections as long as the bulk gap remains open.

\paragraph*{Parity-even magneto-elastic action.} The parity-even functional $S_{\rm even}$ generates the elastic energy and higher-derivative terms for $\bm u$. To quadratic order one may write
\begin{equation}\label{eq:Seven_u_general}
S_{\rm even}[\bm u] = \int dt\,d^2r\left[ -\frac12 C_{ijkl}\,u_{ij}u_{kl} +\frac{\rho_u}{2}(\partial_t u_i)^2 -\frac12 D_{ijkl}\partial_i\partial_j u_k \partial_i\partial_j u_l +\cdots \right]~,
\end{equation}
with $u_{ij}\equiv \frac12(\partial_i u_j+\partial_j u_i)$. The coefficients admit a controlled expansion
\begin{align}\label{eq:coeff_expansion_1overJ}
\begin{split}
C_{ijkl} = C^{(0)}_{ijkl}+\delta C^{(1/J)}_{ijkl} +\mathcal O(& J^{-2})~, \qquad \rho_u=\rho_u^{(0)}+\delta\rho_u^{(1/J)}+\mathcal O(J^{-2})~, \\
D_{ijkl} &=D^{(0)}_{ijkl}+\delta D^{(1/J)}_{ijkl}+\cdots~,    
\end{split}
\end{align}

A convenient matching procedure for elastic moduli is uniform-strain matching: for a constant strain tensor $\varepsilon_{ij}$, define
$\hat{\bm n}_\varepsilon(\bm r)\equiv \hat{\bm n}_0\big((\mathbf 1-\bm\vareps)\bm r\big)$ and expand the zero-temperature effective energy,
\begin{equation}\label{eq:app_elastic_matching}
E_{\rm eff}[\hat{\bm n}_\varepsilon] = E_{\rm eff}[\hat{\bm n}_0] +\frac{1}{2}C_{ijkl}\,\varepsilon_{ij}\varepsilon_{kl}+\cdots~, \qquad
C_{ijkl}\equiv \left.\frac{\partial^2 E_{\rm eff}}{\partial\varepsilon_{ij}\partial\varepsilon_{kl}}\right|_{\varepsilon=0}~.
\end{equation}
The $1/J$ correction to the energy functional follows from the cumulant expansion evaluated on a static background:
\begin{align}\label{eq:app_deltaE_1overJ}
\delta E_{\rm eff}^{(1/J)}[\hat{\bm n}]
&\equiv
\frac{1}{\beta A}\,\delta S_{\rm even}^{(1/J)}[A_\mu=0,\hat{\bm n}]
\nonumber\\
&=
\frac{1}{A}\big\langle \delta H_{1/J}[\hat{\bm n}] \big\rangle_{0}-\frac{1}{2A}\int_0^\beta d\tau\;\big\langle T_\tau\,\delta H_{1/J}(\tau)\,\delta H_{1/J}(0)\big\rangle_{0}^{c} +\cdots~,
\end{align}
and therefore
\begin{equation}\label{eq:app_deltaC}
\delta C^{(1/J)}_{ijkl}=\left.\frac{\partial^2}{\partial\varepsilon_{ij}\partial\varepsilon_{kl}}\delta E_{\rm eff}^{(1/J)}[\hat{\bm n}_\varepsilon]\right|_{\varepsilon=0}~.
\end{equation}
The same logic applies to inertial and higher-gradient coefficients: their $1/J$ corrections are obtained from the small-frequency and small-momentum expansion of the quadratic kernel for $\bm u$ generated by $S_{\rm even}$, with insertions of $\delta\mathcal H_{1/J}$ organized by
Eqs.~\eqref{eq:app_cumulant}--\eqref{eq:app_cumulant_time}.

\paragraph*{Resulting magneto-elastic effective field theory.} Collecting the universal Berry phase term and the parity-even action~\eqref{eq:Seven_u_general}, the long-wavelength skyrmion-crystal effective field theory takes the form
\begin{align}\label{eq:SkX_EFT_with_1overJ}
\begin{split}
S_{\rm SkX}[\bm u;A] = \int dt\,d^2r \,\Big[\frac{\Gamma}{2}\varepsilon_{ij}u_i\partial_t u_j -\frac12 C_{ijkl}u_{ij}u_{kl} +\frac{\rho_u}{2}(\partial_t u_i)^2 +\kappa e A_\mu J^\mu_{\rm sk}[\hat{\bm n}_0(\bm r-\bm u)]\Big] +\cdots~,
\end{split}
\end{align}
where $\Gamma=\kappa Q/A_M$ is fixed by the quantized Hall response, while the parity-even coefficients $(C_{ijkl},\rho_u,\ldots)$ receive controlled non-adiabatic renormalizations starting at order $1/J$.

\section{Landau Level with a Periodic Potential}
\label{app:LL_periodic_potential}

In this Appendix we consider standard formulas for a charged particle in a uniform magnetic field subject to a weak spatially periodic potential. The Hamiltonian is given by
\begin{equation}
H=\frac{(\mathbf{p}-\mathbf{A})^2}{2m}+V(\r)~,
\end{equation}
with uniform magnetic field $B=\varepsilon^{ij}\partial_i A_j$ and cyclotron frequency
$\omega_c=B/m$. The periodic potential is expanded in reciprocal lattice vectors,
\begin{equation}
V(\r)=\sum_{\bm G} V_{\bm G}\,e^{i\bm G\cdot\r}~,
\end{equation}
where $\bm G$ runs over the reciprocal lattice of the potential. To suppress Landau level mixing, we assume the strength of the periodic potential is small compared to the cyclotron energy $|V_{\bm G}|/\omega_c\ll 1$.

Let $\bm a_{1,2}$ be primitive real-space lattice vectors of the potential, so that $V(\r+\alpha_i\bm a_i)=V(\r)$ for integers $\alpha_i$. The corresponding reciprocal lattice vectors $\bm G_{1,2}$ satisfy
\begin{equation}
\bm a_i\cdot\bm G_j=2\pi\,\delta_{ij}~, \qquad \bm G=\alpha_1\bm G_1+\alpha_2\bm G_2~.
\end{equation}
We focus on the special case in which the unit-cell area
\begin{equation}
A_{\rm uc}\equiv|\bm a_1\times \bm a_2|
\end{equation}
encloses exactly one magnetic flux quantum,
\begin{equation}
B A_{\rm uc}=2\pi \qquad \Longleftrightarrow \qquad A_{\rm uc}=2\pi\ell_B^2~, \qquad \ell_B^2 \equiv \frac{1}{B}~.
\end{equation}

Decompose the position operator as
\begin{equation}
r_i=R_i+\bar R_i~,
\end{equation}
where the guiding-center $R_i$ and cyclotron coordinates (or ``Landau orbit'') $\bar{R}_i$ are
\begin{equation}
R_i\equiv r_i-\ell_B^2\varepsilon^{ij}\pi_j~,\qquad
\bar R_i\equiv \ell_B^2\varepsilon^{ij}\pi_j~,\qquad
\pi_i\equiv -i\partial_i-A_i~.
\end{equation}
They obey
\begin{equation}
[R_i,R_j]=-i\ell_B^2\,\varepsilon_{ij}~,\qquad
[\bar R_i,\bar R_j]=+i\ell_B^2\,\varepsilon_{ij}~,\qquad
[R_i,\bar R_j]=0~.
\end{equation}

Define magnetic translation operators acting on the guiding center by
\begin{equation}
T_{\bm q}\equiv e^{i\bm q\cdot \bm R}~.
\end{equation}
Using $[R_x,R_y]=-i\ell_B^2$, they satisfy the projective noncommutative algebra
\begin{equation}
T_{\bm q_1}T_{\bm q_2}
=e^{\,i\ell_B^2 (\bm q_1\wedge\bm q_2)}\,T_{\bm q_2}T_{\bm q_1}
=e^{\frac{i}{2}\ell_B^2 (\bm q_1\wedge\bm q_2)}\,T_{\bm q_1+\bm q_2}~.
\end{equation}
When $B A_{\rm uc}=2\pi$, the primitive operators $T_{\bm G_1}$ and $T_{\bm G_2}$ commute, so one can choose simultaneous eigenstates of $H_0=\frac{(\p-\mathbf{A})^2}{2m}$ and
$\{T_{\bm G_1},T_{\bm G_2}\}$. We denote these magnetic Bloch states by $|n,\bm k\rangle$, where $n=0,1,2,\dots$ is the Landau-level index and $\bm k$ is a magnetic Bloch momentum defined modulo the reciprocal lattice. 

A convenient convention for the action of magnetic translations is
\begin{equation}
T_{\bm q}\,|n,\bm k\rangle
=e^{\frac{i}{2}\ell_B^2 (\bm q\wedge\bm k)}\,|n,\bm k+\bm q\rangle~.
\end{equation}
Because $\bm k$ is defined modulo $\bm G_{1,2}$, these states satisfy quasi-periodic boundary conditions under $\bm k\to \bm k+\bm G_i$. In the one-flux-quantum case this introduces a gauge-dependent sign factor $\eta_{\bm G}$ discussed below.

The cyclotron coordinates can be written in terms of ladder operators
\begin{equation}
\bar R_x=\frac{\ell_B}{\sqrt2}\,(\bar a^\dagger+\bar a)~,\qquad
\bar R_y=\frac{\ell_B}{\sqrt2\,i}\,(\bar a^\dagger-\bar a)~,\qquad
\bar a^\dagger|n,\bm k\rangle=\sqrt{n+1}\,|n+1,\bm k\rangle~.
\end{equation}
Using $e^{i\bm G\cdot\r}=e^{i\bm G\cdot \bm R}\,e^{i\bm G\cdot \bar{\bm R}}$ and the magnetic Bloch boundary conditions, one finds the matrix element
\begin{equation}\label{matrix_compact}
\langle n_1,\bm k_1| e^{i\bm G\cdot\bm r} |n_2,\bm k_2\rangle
=\eta_{\bm G}\delta_{\bm k_1,\bm k_2}\,
e^{-i\ell_B^2(\bm G\wedge\bm k)} e^{-\ell_B^2|\bm G|^2/4}\,
F_{n_1 n_2}(\bm G)~,
\end{equation}
with the form factor
\begin{equation}\label{Fn1n2}
F_{n_1 n_2}(\bm G)=
\sqrt{\frac{n_{\min}!}{n_{\max}!}}\;
L_{n_{\min}}^{\alpha}\!\left(\frac{\ell_B^2|\bm G|^2}{2}\right)
\begin{cases}
\left(\dfrac{-i\ell_B\bar z}{\sqrt2}\right)^{\alpha}, & n_2>n_1~,\\
\left(\dfrac{i\ell_B z}{\sqrt2}\right)^{\alpha}, & n_1\ge n_2~,
\end{cases}
\end{equation}
where $\alpha=|n_2-n_1|$, $n_{\max}=\max(n_1,n_2)$, $n_{\min}=\min(n_1,n_2)$,
$L_a^{b}(x)$ is an associated Laguerre polynomial, and $z\equiv G_x+iG_y$ with $\bar z=G_x-iG_y$. The factor $\eta_{\bm G}=\pm1$ depends on the choice of magnetic Bloch gauge; one convenient convention for one flux quantum per unit cell is
\begin{equation}
\eta_{\bm G}=(-1)^{\alpha_1+\alpha_2+\alpha_1\alpha_2}\qquad
\text{for}\qquad \bm G=\alpha_1\bm G_1+\alpha_2\bm G_2~,
\end{equation}
equivalently $\eta_{\bm G}=+1$ if $\bm G/2$ is a reciprocal lattice vector and $\eta_{\bm G}=-1$ otherwise.

Neglecting Landau-level mixing $|V_{\bm G}|\ll \omega_c$, the Hamiltonian is diagonal in $\bm k$ within each Landau level and acquires the dispersion
\begin{equation}
\varepsilon_n(\bm k) =\omega_c\left(n+\frac12\right)+\tilde\varepsilon_n(\bm k)~,
\qquad \tilde\varepsilon_n(\bm k) =\sum_{\bm G} \langle n,\bm k|V_{\bm G}e^{i\bm G\cdot\r}|n,\bm k\rangle~.
\end{equation}
 with $n_1=n_2=n$ gives
\begin{equation}
\label{eq:supp_disper}
\tilde{\varepsilon}_n(\bm{k})
=\sum_{\bm G} V_{\bm G}\,\eta_{\bm G}\,
e^{-i\ell_B^2(\bm G\wedge \bm k)}\,
e^{-\ell_B^2|\bm G|^2/4}\,
L_n\!\left(\frac{\ell_B^2|\bm G|^2}{2}\right)~,
\end{equation}
where $L_n(x)\equiv L_n^{0}(x)$ is the ordinary Laguerre polynomial.

\bibliography{tmd}

\end{document}